\documentclass[aps,twocolumn,showpacs,preprintnumbers,amsmath,amssymb,amsfonts,superscriptaddress,openany,nofootinbib]{revtex4-1}
\pdfoutput=1 

\usepackage[utf8]{inputenc}
\usepackage{amsfonts}
\usepackage{amsbsy}
\usepackage{url}
\usepackage{enumerate}
\usepackage[justification=raggedright]{caption}
\usepackage{hhline}
\usepackage{graphicx,color}
\usepackage{amssymb,amsmath}
\usepackage{slashed}
\usepackage{hyperref}
\usepackage{braket}
\usepackage{simplewick}
\usepackage{ascmac}
\usepackage{latexsym}
\usepackage{pifont}          
\usepackage{bm}
\usepackage{comment}
\usepackage[normalem]{ulem}
\usepackage{CJKutf8}
\usepackage{cancel}

\graphicspath{{./fig}{./}}

\newcommand{\mr}[1]{\mathrm{#1}}

\newcommand{\f}[2]{\frac{#1}{#2}}

\begin{document}

\preprint{YGHP-24-06, DESY-24-104}

\title{
Tying knots in particle physics
}

\author{Minoru~Eto}
\affiliation{Department of Physics, Yamagata University, Kojirakawa-machi 1-4-12, Yamagata, Yamagata 990-8560, Japan}
\affiliation{Research and Education Center for Natural Sciences, Keio University, 4-1-1 Hiyoshi, Yokohama, Kanagawa 223-8521, Japan}
\affiliation{
International Institute for Sustainability with Knotted Chiral Meta Matter(SKCM$^2$), Hiroshima University, 1-3-2 Kagamiyama, Higashi-Hiroshima, Hiroshima 739-8511, Japan
}

\author{Yu~Hamada}
\affiliation{Deutsches Elektronen-Synchrotron DESY, Notkestr. 85, 22607 Hamburg, Germany}
\affiliation{Research and Education Center for Natural Sciences, Keio University, 4-1-1 Hiyoshi, Yokohama, Kanagawa 223-8521, Japan}

\author{Muneto~Nitta} 
\affiliation{Department of Physics, Keio University, 4-1-1 Hiyoshi, Kanagawa 223-8521, Japan}
\affiliation{Research and Education Center for Natural Sciences, Keio University, 4-1-1 Hiyoshi, Yokohama, Kanagawa 223-8521, Japan}
\affiliation{
International Institute for Sustainability with Knotted Chiral Meta Matter(SKCM$^2$), Hiroshima University, 1-3-2 Kagamiyama, Higashi-Hiroshima, Hiroshima 739-8511, Japan
}

\begin{abstract}
Lord Kelvin's pioneering hypothesis that the identity of atoms is knots of vortices of the aether had a profound impact on the fields of mathematics and physics despite being subsequently refuted by experiments. 
While knot-like excitations emerge in various systems of condensed matter physics,
the fundamental constituents of matter have been revealed to be elementary particles such as electrons and quarks, seemingly leaving no room for the appearance of knots in particle physics.
Here, we show that knots indeed appear as meta-stable solitons in a realistic extension of the standard model of particle physics that provides the QCD axion and right-handed neutrinos.
This result suggests that during the early Universe, a ``knot dominated era'' may have existed, where knots were a dominant component of the Universe,
and this scenario can be tested through gravitational wave observations.
Furthermore, we propose that the end of this era involves the collapse of the knots via quantum tunneling, leading to the generation of matter-antimatter asymmetry in the Universe. 
Our findings exhibit the significant role of knots in particle physics
and represent a modern version of Kelvin's hypothesis.
\end{abstract}

\maketitle


\section*{Introduction}

Knots, mathematically defined as closed curves embedded into three-dimensional space,
appear not only when tying a tie but also in various scientific fields today,
as pioneered by William Thomson, also known as Lord Kelvin.
He conjectured that atoms, thought of as fundamental building blocks of matter at that time, are made from knots of vortices of the aether,
which means that knots are the ultimate origins of all matter in our Universe 
\cite{Thomson:1869}.
Although this elegant conjecture had attracted much attention from many physicists and mathematicians, 
it was eventually falsified by Michelson and Morley's experiment ruling out the existence of aether,
leading to the development of knot theory~\cite{tait1898knots} as a separate mathematical discipline.

Even though Kelvin's idea failed to explain the most fundamental aspects of nature,
the concept of knots has found extensive applications in various domains.
One of the characteristics of knots is their topological nature;
knots do not disappear unless the strings break or intersect,
which has played a crucial role in persistent phenomena \cite{Kauffman:1991ds}.
For instance, knots appear 
as vortex knots 
or topological solitons called Hopfions 
\cite{Faddeev:1996zj}
in diverse subjects such as 
fluid dynamics 
\cite{Moffatt_1969,10.1063/1.881574,Ricca:2009,Kleckner:2013,Arnold:2021}, 
superconductors 
\cite{Babaev:2001zy,Rybakov:2018ktd}, 
Bose-Einstein condensates
\cite{PhysRevE.85.036306,Kleckner2016,
Kawaguchi:2008xi,Hall:2016,Ollikainen:2019dyh},
$^3$He superfluids \cite{Volovik:1977,Volovik:2003fe}, 
nematic liquid crystals
\cite{PhysRevLett.110.237801,
PhysRevLett.113.027801,
Ackerman:2015,Ackerman:2017,
Ackerman:2017b,
Tai:2018,Tai:2019,RevModPhys.84.497,Smalyukh:2020zin,Smalyukh:2022},
colloids, 
magnets \cite{Kent:2020jvm}, 
optics \cite{Dennis2010}, 
electromagnetism \cite{Trautman:1977im,Ranada:1989wc,
Kedia:2013bw,Arrayas:2017sfq},
and active matter \cite{Shankar2022}.
The stability of the solitons is ensured by the topological invariant associated with knots.

In modern physics, the understanding of the fundamental nature of matter has progressed from atoms to nucleons and ultimately to quarks and gluons
primarily through elementary particle physics,
instead of knots.
Here a natural question may arise: Are knots truly irrelevant within the realm of elementary particle physics?
It is known that a knot invariant in mathematics arises from
topological field theories known as the Chern-Simons theory~\cite{Witten:1988hf},
which exhibits a direct connection between theoretical physics and mathematics.
In addition, a knot as a soliton is proposed~\cite{Faddeev:1996zj,Battye:1998pe,Manton:2004tk,Radu:2008pp,Kobayashi:2013xoa,Shnir:2018yzp} in a field-theoretic toy model called the Faddeev-Skyrme model,
in which the stability of the knot soliton requires non-renormalizable interactions
destroying the predictability of the model.
Nevertheless, knots have received relatively less attention compared to condensed matter physics and other fields.
In particular, knots have never appeared in phenomenologically viable models in the context of particle physics.

Here we revisit Kelvin's idea in elementary particle physics.
We theoretically show the existence of stable knot solitons in a particle physics model for the first time.
The model is a simple setup 
containing global and local $U(1)$ symmetries denoted by $U(1)_\mr{global}$ and $U(1)_\mr{local}$, respectively.
The two symmetries are spontaneously broken by condensations of two complex scalar fields, 
and thus a superfluid 
and superconductor are simultaneously realized,  
resulting in 
the production of two kinds of string-like defects called 
flux tubes and superfluid vortices, respectively. 
String loops (either knots or unknots) made of the single species of the flux tube or superfluid vortex shrink  due to their tensions and eventually disappear 
\cite{PhysRevE.85.036306,Kleckner2016}. 
In contrast, it turns out that the knot solitons made by linking the two types of the strings remain stable.
Such configurations are known as two-component links in knot theory.
The stability is realized by the so-called Chern-Simons coupling between the Nambu-Goldstone (NG) boson and the massive $U(1)_\mr{local}$ gauge field.
The gradient of the NG field and the magnetic flux of the gauge field induce a $U(1)_\mr{local}$ electric charge and hence electric flux around the string loop.\footnote{
Note that the stability mechanism by the Chern-Simon coupling 
is crucial, 
although 
knots were discussed 
in similar systems 
without such a coupling: 
liquid metalic hydrogen 
\cite{Babaev:2004rm,Smiseth:2004na} 
and 
the coexistence of 
a neutron superfluid 
and a proton superconductor 
inside neutron star interior 
\cite{Babaev:2002wa}.
}

\begin{figure*}[tbp]
\centering
\includegraphics[width=0.8\textwidth]{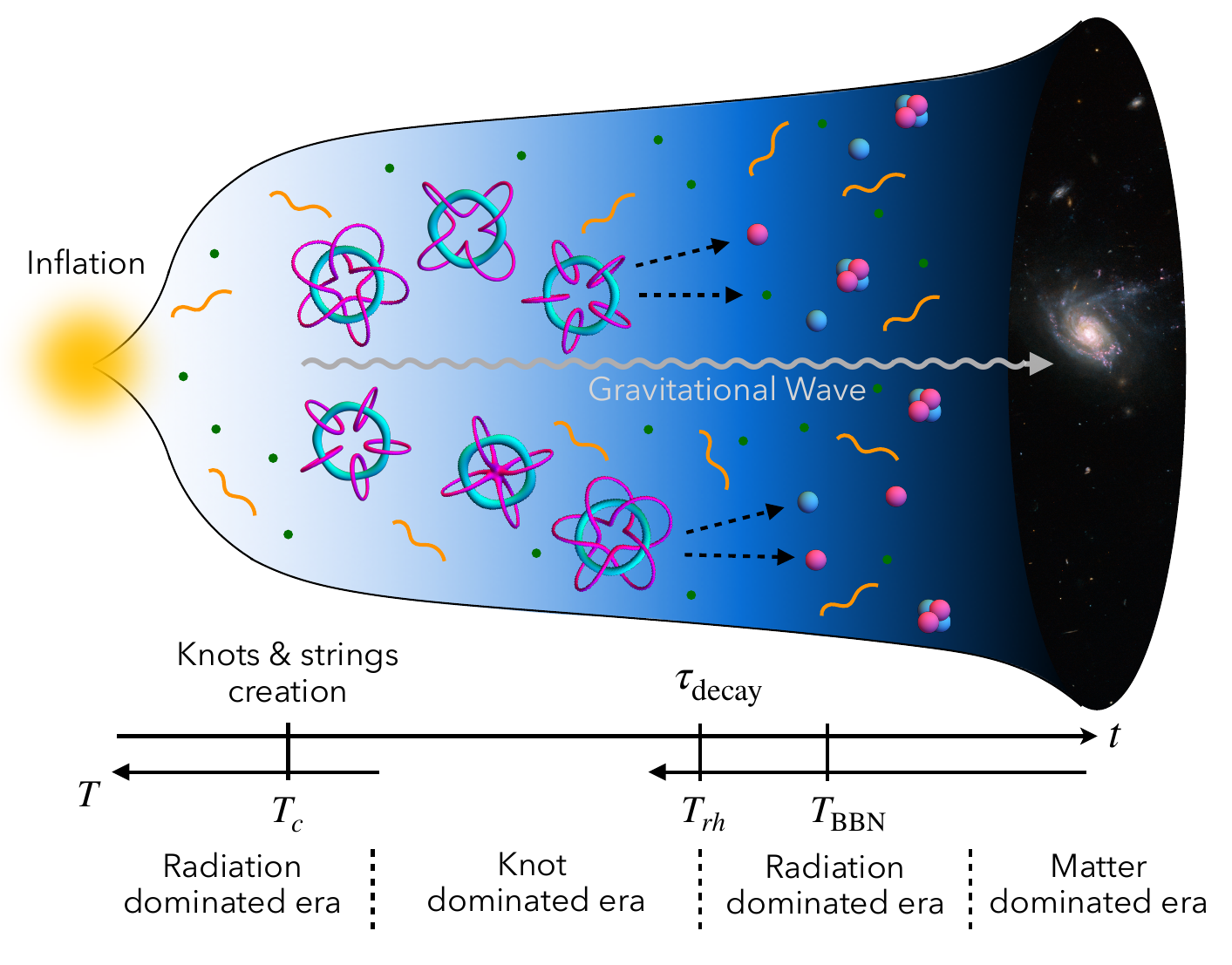}
\caption{
Illustration of cosmic history in our scenario.
After creation of the strings and knots, 
the knots eventually dominate the energy density of the Universe (knot dominated era).
At $t\sim \tau_\mathrm{decay}$, they decay into light particles, leading to radiation dominated era again.
This decay can produce the correct baryon asymmetry in the Universe via the non-thermal leptogenesis.
In addition, the existence of the knot domination is imprinted on stochastic GW spectrum radiated from the string network,
which can be tested by future GW observations.
}
\label{fig:history}
\end{figure*}

By identifying the $U(1)_\mr{global}$ and $U(1)_\mr{local}$ symmetries with
the $U(1)$ Peccei-Quinn symmetry \cite{Peccei:1977hh,Peccei:1977ur} and
$U(1)$ $B-L$ (baryon number minus lepton number) symmetry,
which are popular hypothetical symmetries to resolve mysteries unanswered by the Standard Model (SM) of particle physics,
our setup becomes a realistic model as a microscopic theory beyond the SM.
In addition, we below argue a cosmological scenario that the knot solitons were produced at the early stage of the Universe.
Because they can be sufficiently abundant to dominate the energy density of the Universe,
this scenario is testable in terms of stochastic signals of gravitational waves. 
Furthermore, the knots eventually decay by a quantum effect,
and the decay can produce 
an imbalance between matter and antimatter,
which explains why matter is dominant compared to antimatter in the present Universe, 
see Fig.~\ref{fig:history}.
Thus, this scenario is a modern revival of Kelvin's dream; 
The Universe was full of knots, and matter originated from them.

\section*{The model}
We start with a model that includes two complex scalar fields $\phi_1$ and $\phi_2$ and a $U(1)$ gauge field.
The Lagrangian for the scalars is given as
\begin{equation}
 \mathcal{L}_{\phi} = |D_\mu \phi_1|^2 + |D_\mu \phi_2|^2 - V(\phi_1,\phi_2) 
\label{eq:scalar-lagrangian}
\end{equation}
with the potential term\footnote{We here assume an exchange symmetry between the two scalars in the potential for simplicity. This is however not necessary in the following argument; we may add other terms such as $|\phi_2|^4$.
}
\begin{equation}
V = \lambda(|\phi_1|^2 + |\phi_2|^2 - m^2)^2  - \kappa |\phi_1|^2 |\phi_2|^2 \, ,
\end{equation}
and $\lambda, \kappa>0$.
The Lagrangian is invariant under each $U(1)$ phase rotation of the scalars $\phi_1$ and $\phi_2$, 
which we denote $U(1)_1$ and $U(1)_2$, respectively.
The covariant derivative is given as $ D_\mu \phi_{1(2)} \equiv (\partial _\mu - i q_{1(2)} g A_\mu)\phi_{1(2)}$,
where $A_\mu$ is the $U(1)$ gauge field and $g$ is its coupling constant.
$q_1$ and $q_2$ denote $U(1)$ gauge charges of $\phi_1$ and $\phi_2$, respectively.
For simplicity, we take $q_1=1$ and $q_2=0$ for the moment,
leading to that
$U(1)_1$ is promoted to a gauge symmetry called $U(1)_\mr{local}$ 
while $U(1)_2$ is a global symmetry denoted by $U(1)_\mr{global}$.
The Lagrangian for the kinetic term of $A_\mu$ is given as $\mathcal{L}_g = (-1/4) F_{\mu\nu} F^{\mu\nu}$ with $F_{\mu\nu}$ being the field strength.
Thus the full Lagrangian is given as $ \mathcal{L} = \mathcal{L}_\phi + \mathcal{L}_g$.

Because of the tachyonic mass parameter $-\lambda m^2<0$ in $V$, 
the two scalars condense and develop vacuum expectation values (VEVs) as
$ \langle \phi_1 \rangle=\langle \phi_2 \rangle = v/\sqrt{2}$,
which spontaneously break both of the $U(1)_\mr{local}$ and $U(1)_\mr{global}$ symmetries.

At the broken phase,  
quantum corrections can induce the so-called Chern-Simons coupling in addition to the Lagrangian $\mathcal{L}_\phi + \mathcal{L}_g$,
which characterizes the coupling between the NG boson $a\equiv -i \, \mathrm{arg} \,\phi_2$ and $A_\mu$, given as
\begin{equation}
 \mathcal{L}_\mr{CS} = C \,a  F_{\mu\nu}  \tilde F^{\mu\nu}, \label{eq:CS-coupling}
\end{equation}
where $\tilde F ^{\mu\nu}= \epsilon^{\mu\nu\lambda\rho} F_{\lambda\rho}$
and $C$ is a model-dependent dimensionless constant; for instance, it may be induced by one-loop effects of fermions.\footnote{
If we assume the existence of a UV theory for our model, 
the coupling $C$ is quantized as
$C= N g^2/32\pi^2$ with an integer $N$.
In such a case, this model possesses 
an emergent symmetry at low energy, 
characterized by 
a mathematical structure called 
the 4-group  \cite{Hidaka:2021mml,Hidaka:2021kkf}
while it is the 3-group 
in the unbroken phase of the $U(1)_\mr{local}$ symmetry
\cite{Hidaka:2020izy,Hidaka:2020iaz}.
}
To consider general setups,
$C$ is taken as a free parameter in this paper.
The model $\mathcal{L}_\phi + \mathcal{L}_g + \mathcal{L}_{\rm CS}$ includes
the axion electrodynamics \cite{Wilczek:1987mv}, 
which has been recently applied 
in condensed matter physics to describe topological superconductors \cite{Qi:2012cs,Stone:2016pof,Stalhammar:2021tcq}.

The spontaneous breaking of the two symmetries $U(1)_\mr{local}$ and $U(1)_\mr{global}$ (or, $U(1)_1$ and $U(1)_2$)
gives rise to two kinds of vortex strings as topological defects.
We call these two strings the $\phi_1$ and $\phi_2$ strings;
for the $\phi_1$ string, the phase of $\phi_1$ winds around the defect while for the $\phi_2$ string the phase of $\phi_2$ winds,
both of whose windings are topologically protected.
Since the $\phi_1$ string is a local string, it contains a magnetic flux associated with $A_\mu$, $|\Phi_A| = 2\pi/g$,
like flux tubes in superconductors
while the $\phi_2$ string is a global string without magnetic fluxes like superfluid vortices.

\section*{Knot soliton}

\begin{figure*}[tbp]
\centering
\includegraphics[trim = 55 55 55 55, width=0.3\textwidth]{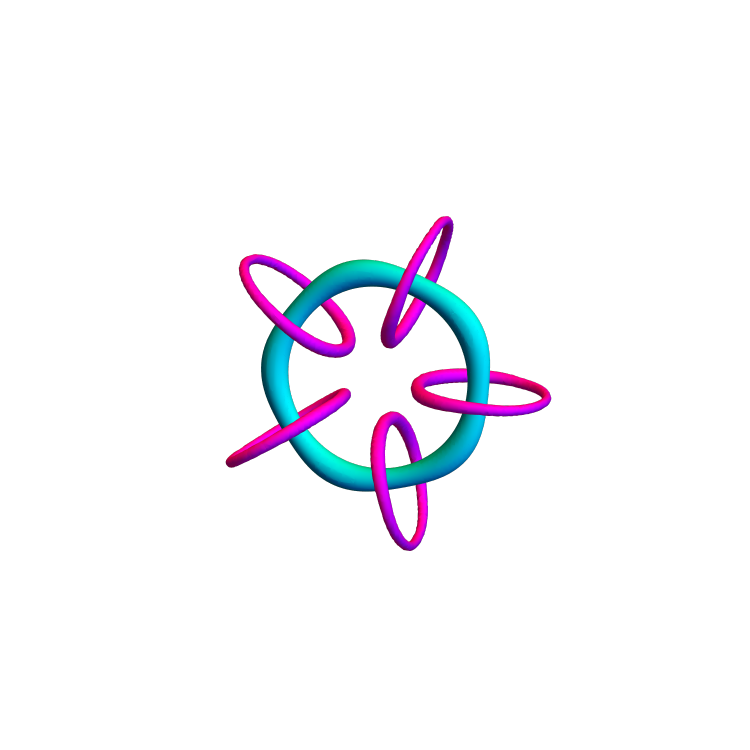} \hspace{0mm}
\includegraphics[trim = 55 55 55 55, width=0.3\textwidth]{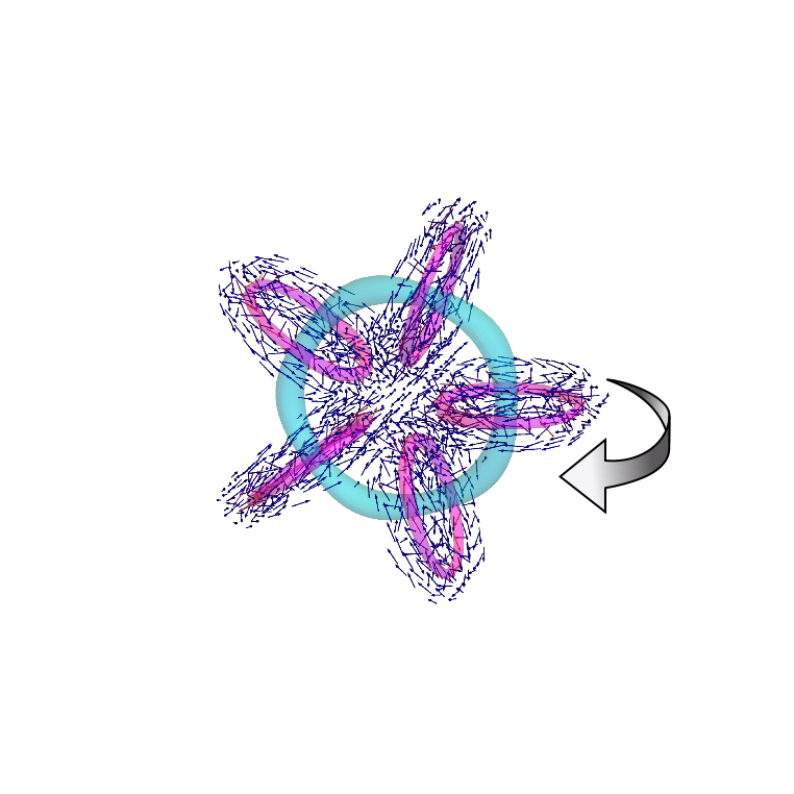} \hspace{0mm}
\includegraphics[trim = 55 55 55 55, width=0.3\textwidth]{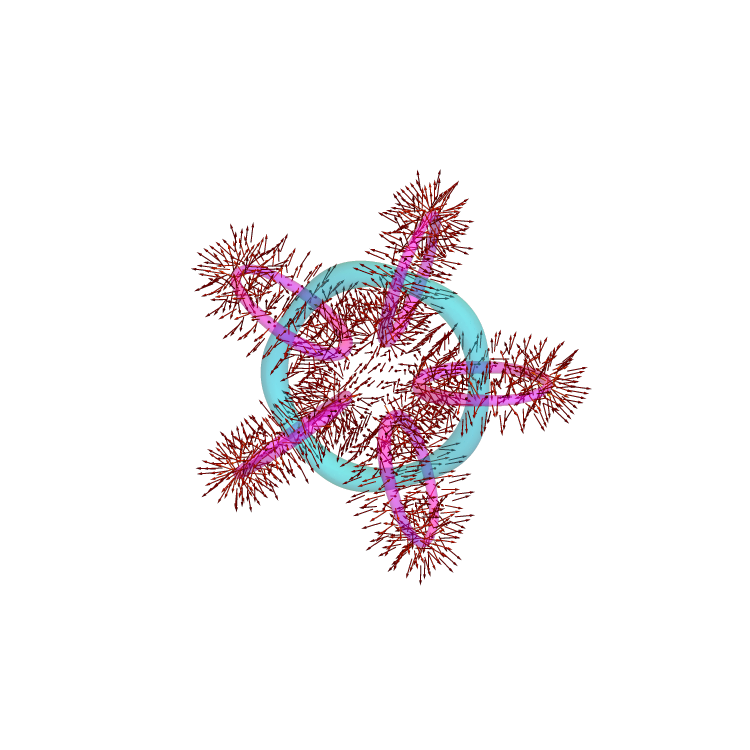} 
\caption{
3D plots of numerical solution for a knot soliton.
The magenta and cyan regions represent the cores of the $\phi_1$ and $\phi_2$ strings ($|\phi_1|^2/v^2<0.1$ and $|\phi_2|^2/v^2<0.1$).
The single $\phi_2$ string loop is linked with five $\phi_1$ string loops,
so that the solution has the linking number $5$.
In the middle and right panels, the arrows that are overlaid with transparent string cores
indicate the magnetic flux $\vec B$ and the electric flux $\vec E$ of the massive gauge field $A_\mu$, respectively,
and their colors correspond to their strength $|\vec{B}|$ and $|\vec{E}|$.
The magnetic flux is along the $\phi_1$ strings, one of whose direction is indicated by the large gray arrow,
 while the electric one comes out from the strings,
which means that the strings have net electric charges.
We here take $\lambda/g^2=10^3$, $\kappa/g^2 = 0.0008$, and $C=400$.
The energy of the solution is $7.0\times 10^3 \,v/g$.
}
\label{fig:link}
\end{figure*}

\begin{figure*}[tbp]
\centering
\includegraphics[trim = 55 55 55 55, width=0.23\textwidth]{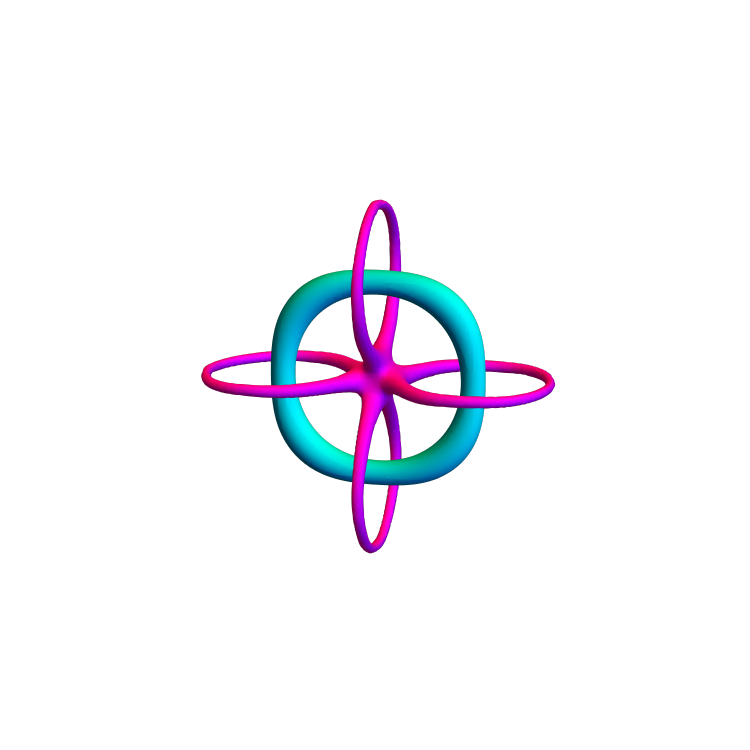} \hspace{1ex}
\includegraphics[trim = 55 55 55 55, width=0.23\textwidth]{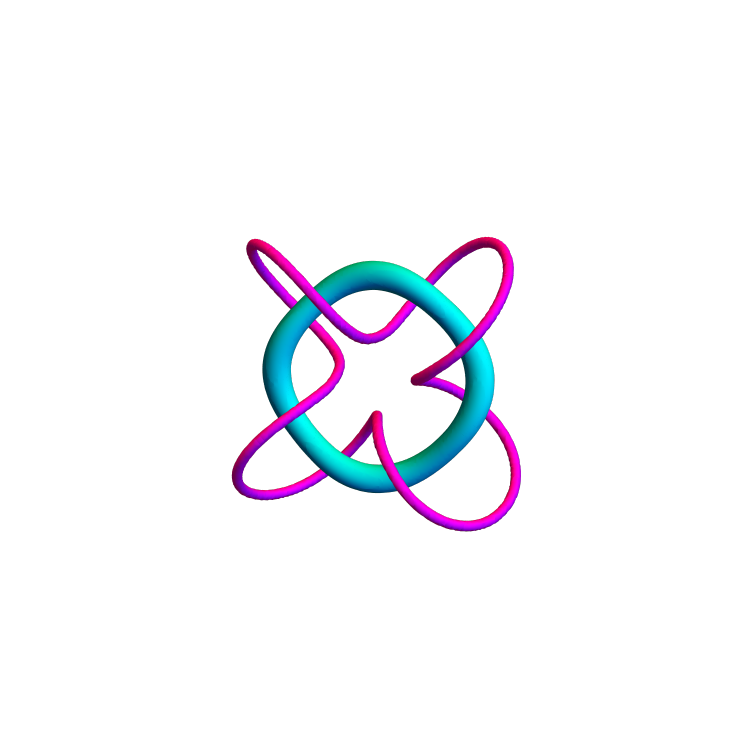} \hspace{1ex}
\includegraphics[trim = 55 55 55 55, width=0.23\textwidth]{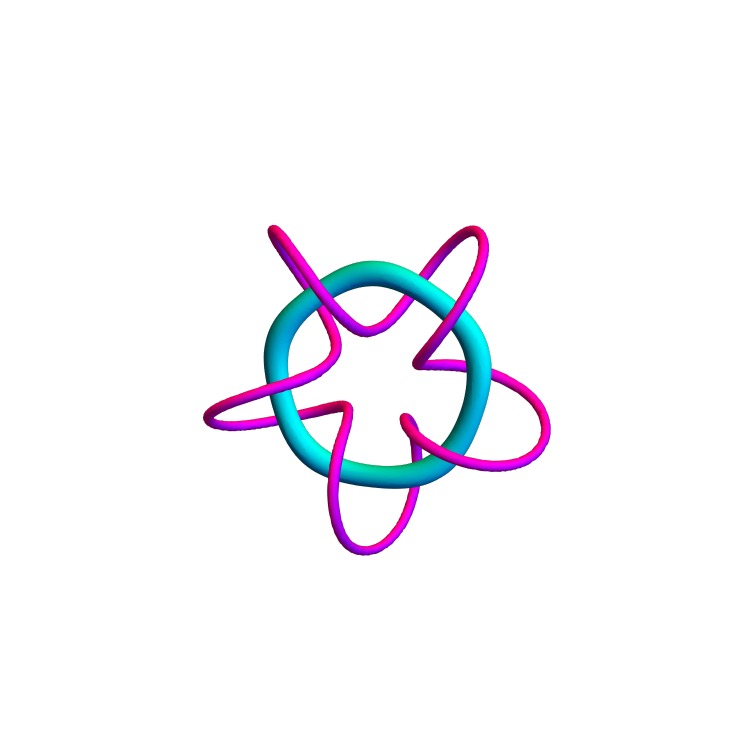} \hspace{1ex}
\includegraphics[trim = 55 55 55 55, width=0.23\textwidth]{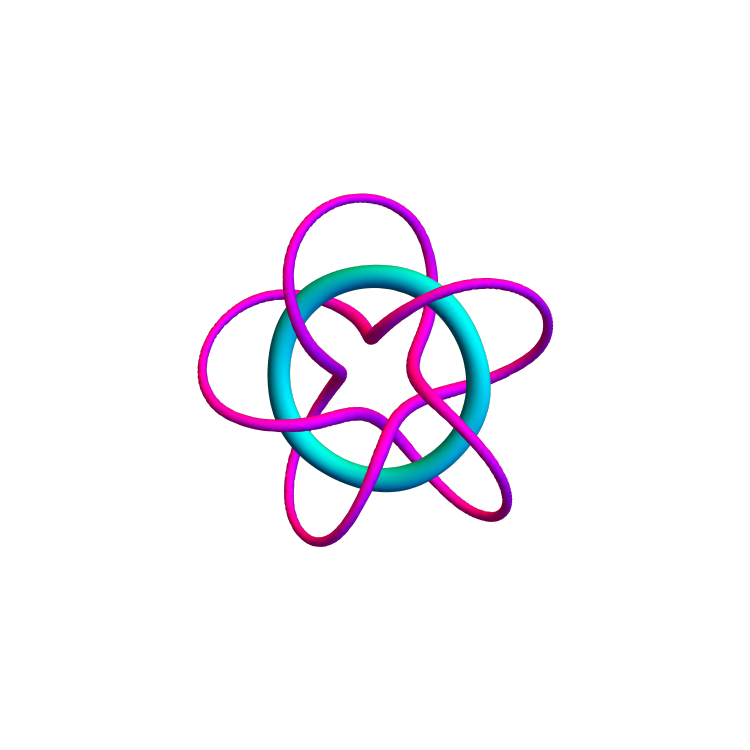} 
\caption{
3D plots of numerical solutions for other knot solitons.
The colors and the parameters are the same as those in Fig.~\ref{fig:link}.
The two left plots have the same linking number $4$
and the the other plots have the linking number $5$.
The energy of the solutions are $6.0\times 10^3 \,v/g$, $6.3 \times10^3 \,v/g$, $7.3\times 10^3 \,v/g$ and $7.5\times 10^3 \,v/g$ (from left to right).
}
\label{fig:other-sols}
\end{figure*}

The Chern-Simons coupling~\eqref{eq:CS-coupling} does not affect the individual $\phi_1$ or $\phi_2$ string
because the $\phi_1$ ($\phi_2$) string does not contain excitation of the NG boson $a$ (the gauge field $A_\mu$), resulting in that \eqref{eq:CS-coupling} vanishes.
Thus a loop made of either the $\phi_1$ or $\phi_2$ string is not stable but shrinks due to their tension and eventually disappears.

Now, make the $\phi_1$ and $\phi_2$ strings linked.
Particularly, one simple example is a link made of a single $\phi_2$ string and multiple $\phi_1$ strings (See the left panel of Fig.~\ref{fig:link}).
Now the coupling \eqref{eq:CS-coupling} plays a crucial role;
it induces the electric flux of the $A_\mu$ field around the $\phi_1$ string.
To see this, let us take a functional derivative of the Chern-Simons coupling with respect to $A_0$ as
\begin{align}
\frac{\delta }{\delta A_0} \int d^4x \, \mathcal{L}_\mr{CS}
& = - C  \, \left(\vec {\nabla} a \cdot \vec B \right) \, , \label{eq:CS}
\end{align}
where $\vec{B}$ is the magnetic field of $A_\mu$, $\vec{B}=\nabla \times\vec{A}$, and we have used that all fields are static.
Eq.~\eqref{eq:CS} changes the Gauss law in the Maxwell equation with respect to $A_0$ and hence $C \, \vec{\nabla} a \cdot \vec{B}$ behaves as a source of the electric field.
\footnote{
A similar effect is well known for Chern-Simons vortices in $2+1$ dimensions~\cite{Horvathy:2007ps,Horvathy:2008hd}.
}
The total electric charge is estimated to be
$C \int d^3 x \,\vec{\nabla} a \cdot \vec{B} 
= 4\pi^2 \, C N_\mr{link}/ g$
where the strings are assumed to be sufficiently thin~\cite{Yokoi_2013}.
The integer $N_\mr{link}$ is the linking number of the $\phi_1$ and the $\phi_2$ strings.\footnote{Formally this is the same as cross helicity in magnetohydro dynamics when we regard $\vec\nabla a$ as the velocity of fluid~\cite{Yokoi_2013,Nastase:2022aps}.}
From this, the $\phi_1$ strings linking with the $\phi_2$ strings get the electric charge, 
which gives rise to the electric field $\vec E$ around the strings.
Even though this electric field is screened due to the VEV of $\phi_1$ like Meissner effect,
for large enough $C$, the $\phi_1$ string loops are prevented from shrinking by the electric field and is stabilized.\footnote{
A similar vortex loop with the electric charge/current is known as the vorton~\cite{Davis:1988ij,Davis:1988jq}.
The classical and quantum studies of the stability of vortons are given 
in Refs.~\cite{Battye:2008mm,Radu:2008pp,Garaud:2013iba,Battye:2021sji,Battye:2021kbd,Ibe:2021ctf,Abe:2022rrh}.
Our knot soliton considered here is however different from those in the sense that it does not contain internal degrees of freedom on the loop.
}

In order for the linked configuration to be stable under arbitrary perturbations,
another condition $\lambda \gg g^2, \kappa$ is necessary.
If this is met, the link of the 
$\phi_1$ and $\phi_2$
strings cannot be de-linked,
i.e., the strings cannot pass through each other.
This is because
in the limit of $\lambda\to \infty$,
the linking number is equivalent to the Skyrmion number~\cite{Gudnason:2020luj,Gudnason:2020qkd}
which is characterized by the homotopy group $\pi_3 (S^3)\simeq {\mathbb Z}$ defined by a map from $S^3$, 
obtained by a compactification of the real space ${\mathbb R}^3$, 
to $S^3$ defined by $|\phi_1|^2 + |\phi_2|^2 = \mathrm{const.}$, 
and hence in this regime, this number approximately becomes a topological invariant.\footnote{
This is similar to the case of Skyrmions in two-component Bose-Einstein  condensates~\cite{Ruostekoski:2001fc,Battye:2001ec,Nitta:2012hy}.
}
In practical cases with finite $\lambda$, the $\phi_1$ and $\phi_2$ strings are repelled from each other~\cite{Eto:2011wp,Rybakov:2018ktd} as far as the condition $\lambda \gg g^2, \kappa$ is met,
and hence they cannot be overlapped.

We numerically solve the classical equations of motion (EOMs) to confirm the existence of the stable knot soliton in the model.
We perform a minimization procedure for the energy of the configuration starting from an appropriate initial configuration with the Gauss-law constraint.
See Appendix~\ref{app-numerical} for the details.
After the minimization converges with desired accuracy, the configuration is regarded as the static solution of the EOMs.
Fig.~\ref{fig:link} shows the cores of the strings of the obtained solution, in which the magenta and cyan colors represent regions $|\phi_1/v|^2<0.1$ and $|\phi_2/v|^2<0.1$, respectively.
We here take $\lambda/g^2=10^3$, $\kappa/g^2 = 0.0008$, and $C=400$.
The magnetic and electric fields are also shown by arrows.
In this figure, the single $\phi_2$ string loop is linked with five $\phi_1$ string loops (with the same linking direction),
so that the solution has the linking number $5$.
One can also see that the magnetic flux is along the $\phi_1$ strings, one of whose direction is indicated by the large gray arrow.
On the other hand, the electric one comes out from the strings,
which means that the strings have net electric charges.
This is consistent with the argument above.
\footnote{
It is known that the electric flux may cause an instability for the NG boson and gauge field
due to the Chern-Simons coupling~\cite{Ooguri:2011aa}.
Even though our setup has some differences from theirs, such as the mass of the gauge field,
this effect still exists in our case.
Consequently, the NG boson $a$ and gauge field $A_i$ exhibit oscillating profiles inside the $\phi_1$ strings.
However, profiles inside the strings are not crucial for the stability 
because the amount of the electric charge is fixed by the linking number
unless the strings do not have overlap.
}

Note that a similar solution is obtained 
by acting parity transformation on the knot soliton shown here.
This solution has the opposite linking direction of the $\phi_1$ and $\phi_2$ strings compared to the knot soliton, so that we call this the anti-knot soliton.
The anti-knot one has the $U(1)_X$ electric charge (and Skyrme number) with the opposite signs to those of the knot one.

We can also obtain other knot solitons,
which are shown in Fig.~\ref{fig:other-sols}.
The leftmost panel shows a single loop of the $\phi_2$ string linking with four $\phi_1$ string loops.
In the other panels, single $\phi_1$ and $\phi_2$ loops make higher linking number by linking multiple times.
The colors are the same as those in Fig.~\ref{fig:link}.
For both types, the energy grows as the linking number gets higher
because the electric charge contained in the $\phi_1$ string becomes larger.
The energy of all configurations are almost dominated by the $\phi_1$ strings and the electric flux.
Within the same linking number, the energy gets larger as the length of the $\phi_1$ string gets longer.
Therefore, the leftmost solution in Fig.~\ref{fig:other-sols} and that in Fig.~\ref{fig:link}
have the least energy with the linking number $4$ and $5$, respectively.
Although we did not find any more other knot solitons 
with our numerical code, this does not necessarily mean non-existence of other solitons.

Even when the (anti-)knot solitons are classically stable, they can decay due to quantum tunneling,
whose lifetime is denoted by $\tau_\mr{decay}$.
One possible decay process is the de-linking process, i.e., the $\phi_1$ and $\phi_2$ strings pass through each other by quantum tunneling.
The calculation of the decay rate is complicated and beyond the scope of this paper.
Instead, we will later relate the decay rate to the cosmological history of the knot solitons.

\section*{Particle phenomenology}

In spite of the great success of the SM,
there are still several problems that the SM is not able to answer,
such as dark matter of the Universe, tiny masses of the neutrinos, strong CP problem, and baryon asymmetry in the Universe.
(For a review, see, e.g., Ref.~\cite{ParticleDataGroup:2022pth}.)
It is believed that (some of) these problems should be resolved by new physics beyond the SM.

Particularly, one elegant way to extend the SM is to introduce new symmetries that do not exist within the SM.
The strong CP problem and dark matter are simultaneously explained by introducing 
a global $U(1)$ symmetry called the Peccei-Quinn symmetry \cite{Peccei:1977hh,Peccei:1977ur} (denoted by $U(1)_\mathrm{PQ}$), 
which spontaneously breaks and provides a NG boson, the QCD axion~\cite{Weinberg:1977ma,Wilczek:1977pj}.
In addition, it is well motivated to introduce the gauge $B-L$ (baryon number minus lepton number) symmetry, $U(1)_{B-L}$,\footnote{The SM contains \textit{global} $U(1)_{B-L}$ symmetry as an accidental symmetry although it cannot be gauged due to the gauge anomaly in the SM unless right-handed neutrinos are introduced.}
because it requires the existence of the right-handed neutrinos 
and its spontaneous breaking gives them Majorana masses, which naturally explain the tiny neutrino masses in terms of the type-I seesaw mechanism ~\cite{Minkowski:1977sc,Gell-Mann:1979vob,Mohapatra:1979ia,Yanagida:1979as}.
The right-handed neutrinos are also expected to play key roles to generate baryon asymmetry in the Universe.
Furthermore, the $U(1)_{B-L}$ gauge symmetry naturally arises from grand unified theories (GUTs).

Now we apply our study on the knot soliton to particle physics models and discuss its influence on cosmology.
One realistic and promising setup is obtained by identifying the $U(1)_\mathrm{local}$ and $U(1)_\mathrm{global}$ symmetries in \eqref{eq:scalar-lagrangian} as the $U(1)_{B-L}$ and $U(1)_\mathrm{PQ}$ symmetries, respectively.
In order for $\phi_1$ to couple with right-handed neutrinos, 
the $B-L$ charge of $\phi_1$ is taken as $q_1=2$ instead of $1$
while keeping $q_2=0$.
We emphasize that this setup is well motivated because it can elegantly resolve some of the mysteries in particle physics as stated above.\footnote{
When one allows $\phi_2$ to have a small non-zero charge under $U(1)_{B-L}$
and introduces fermionic matters charged under the $U(1)_{B-L}$ and $U(1)_\mathrm{PQ}$ symmetries,
the setup may have a bonus~\cite{Fukuda:2017ylt,Ibe:2018hir} that
the quality of the $U(1)_\mathrm{PQ}$ symmetry is protected against the effect of quantum gravity.
(See Refs.~\cite{Fukuda:2017ylt,Ibe:2018hir} for more details.)
}
In this setup, the QCD axion is identified with the phase of $\phi_2$.
For this to solve the strong CP problem,
we need additional fermionic fields inducing the Chern-Simons coupling between the QCD axion and gluon,
which may be realized by introducing
Kim-Shifman-Vainshtein-Zakharov(KSVZ)-like~\cite{Kim:1979if,Shifman:1979if} heavy quarks.
How the QCD axion couples to the $B-L$ and SM gauge bosons depends on the charge assignment of the fermionic sectors,
which is kept as general as possible hereafter.
From the astrophysical constraints on the QCD axion,
we take $v \gtrsim 10^{8} \, \mathrm{GeV} $~\cite{Irastorza:2018dyq,Chang:2018rso,Hamaguchi:2018oqw,Caputo:2024oqc}.

\section*{Cosmology and gravitational wave}

Now let us discuss cosmological fate of the knot solitons.
The Universe was very hot and full of thermal relativistic particles at early stages (after the cosmic inflation).
When the temperature $T$ of the Universe is much larger than $v$,
the $U(1)_{B-L}$ and $U(1)_\mathrm{PQ}$ symmetries are not broken
while they get spontaneously broken almost simultaneously due to a phase transition when $T$ gets less than a critical temperature 
$T_c \simeq v$.
At the transition, the $\phi_1$ and $\phi_2$ strings are produced by the Kibble-Zurek mechanism~\cite{Kibble:1980mv,Zurek:1985qw}.
The produced strings have random configurations,
in which some of the strings are linked with finite probability.
After the oscillation and the random motion are relaxed, 
long $\phi_1$ and $\phi_2$ strings form conventional networks whose typical length scale is the Hubble scale.
Besides them, 
small loops with the linking configurations become the knot solitons that we have studied above.
Let us estimate the number density of the knot solitons produced by this mechanism.
One should note that, when $\kappa$ is sufficiently small compared to the other parameters $\lambda,g$ and $C$, 
the scalar potential $V$ has an approximate global $O(4)$ symmetry spontaneously broken down to $O(3)$.
Thus, within a microscopic length scale smaller than $1/(\sqrt{\kappa} T_c)$,
the linking number density is well approximated by the number density of textures~\cite{Turok:1989ai,Borrill:1991mv,Leese:1991gt,Vilenkin:2000jqa} rather than that of linking cosmic strings~\cite{Vachaspati:1994ng}.
According to Ref.~\cite{Turok:1989ai},
the linking number density is given as $\sim  0.04 \, \xi^{-3} $, 
where $\xi$ is the correlation length to be $\xi \sim T_c^{-1}$.
Noting that the knot solitons that we found above should have the linking number larger than or equal to $4$,
the number density of the (anti-)knot solitons is estimated as
$ n_\mathrm{knot}(T=T_c) \sim (0.04)^{4} \,\xi^{-3} $.\footnote{Note that 
results given below do not depend on the value of the produced number density $n_\mathrm{knot}(T_c)$ as long as they dominate the energy density of the Universe before they decay.}

After the production, the knot solitons behave as heavy particles decoupled from the thermal bath
with the long lifetime $\tau_\mathrm{decay}$,
and hence their energy density is given as $ \rho_\mathrm{knot}= M\, n_\mathrm{knot}$
with the mass of the knots $M\simeq \mathcal{O}(10^3) \, v/g $.
Since $\rho_\mathrm{knot}$ is diluted by the cosmic expansion being proportional to the cubic of the scale factor $a(t)^{-3}$, 
which is slow compared to the radiation energy density in the thermal plasma, $\rho_\mathrm{rad}\propto a(t)^{-4}$,
the energy density of the knots eventually overcomes the latter and dominates the total energy density of the Universe
when the temperature reaches $T \sim M\, n_\mathrm{knot}(T_c)/(g_* T_c^{3})$.
Here $g_*$ is the number of the degrees of freedom for relativistic particles, $g_*= \mathcal{O}(10^2)$.
Note that this domination persists until they decay.
We call this epoch ``knot dominated era''.
When cosmic time exceeds the lifetime of the knots, $t \sim \tau_\mathrm{decay}$,
they decay by the quantum tunneling
releasing their energy into light particles,
which leads to the radiation dominated epoch again via a secondary reheating.
Assuming instantaneous decays,
the temperature of the reheated thermal bath is found to be~\cite{Kolb:1990vq}
\begin{align}
 T_{rh}& \simeq\left(\frac{30}{\pi^2 g_*}\f{3 M_\mr{pl}^2}{8\pi} \,\Gamma^2\right)^{1/4}  \, .
\end{align}
with $\Gamma \equiv 1/\tau_\mr{decay}$ and $M_\mr{pl}$ being the decay rate and the Planck mass, respectively.
Note that $\Gamma$ is typically exponentially suppressed
because of non-perturbative processes.
In order to realize the successful Big-Bang Nucleosynthesis (BBN), $T_{rh}$ must be larger than $\mathcal{O}(1)$ MeV~\cite{Kawasaki:1999na,Kawasaki:2000en,Hannestad:2004px}.
Therefore, the Universe experiences the knot dominated era 
sandwiched by two radiation dominated eras before BBN.
See Fig.~\ref{fig:history}.

\begin{figure}[tbp]
\centering
\includegraphics[width=0.48\textwidth]{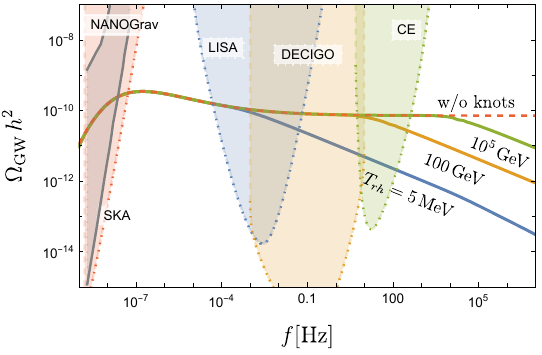} 
\caption{
Stochastic GW spectrum $\Omega_\mathrm{GW} h^2 $ from the network of the $\phi_1$ strings with $G\mu=10^{-11}$,
in which $G$ and $\mu$ are the Newton constant and 
the $\phi_1$ string tension given as $\mu\simeq v^2$.
The red dashed line corresponds to the cosmological evolution without the knot solitons
while the solid ones include the knot dominated era
with different reheating temperatures due to the knot decay,
$T_{rh}=5$~MeV (blue), $100$~GeV (orange), and $10^5$~GeV (green).
We also show the observed signals of NANOGrav 15-year data set~\cite{NANOGrav:2023gor} as gray region
and projected sensitivity curves of future observatories:
DECIGO~\cite{Kawamura:2020pcg} (orange), 
Cosmic Explorer (CE)~\cite{Reitze:2019iox} (green),
LISA~\cite{Bartolo:2016ami} (blue), 
and Square Kilometer Array (SKA)~\cite{Janssen:2014dka} (red). 
}
\label{fig:GW}
\end{figure}

Remarkably, this existence of the knot dominated era
can be observed in terms of gravitational waves (GW)
when there exists a sufficient GW source.
Fortunately, this model already contains a possible source of GW: the string network.
Thermal history of the Universe including the knot dominated era is imprinted on the GW spectrum.
Following the study given in Refs.~\cite{Cui:2017ufi,Cui:2018rwi,Gouttenoire:2019kij}, 
we obtain an expected spectrum of the stochastic GW emitted from the network of the $\phi_1$ strings
depending on $v$ and $T_{rh}$
as shown in Fig.~\ref{fig:GW}.
Here $\Omega_\mathrm{GW} h^2 $ is defined as 
$\Omega_\mathrm{GW} h^2 = (d \rho_\mathrm{GW}(f)/ d \log f) h^2/\rho_c$
with $h$ being the dimensionless Hubble constant, 
$\rho_\mathrm{GW}(f)$ the GW energy density for a frequency $f$ observed today, and
$\rho_c$ the critical energy density.
We also show in Fig.~\ref{fig:GW} the observed signals of NANOGrav 15-year data set~\cite{NANOGrav:2023gor} as gray region\,\footnote{
We do not aim to explain the NANOGrav 15-yrs signals by the GW spectrum from the $\phi_1$ strings 
because the signals are unlikely to be explained by cosmic strings alone~\cite{NANOGrav:2023hvm}.
Instead they may be explained by combining the spectrum of the strings with those of super massive black holes.
}
and projected sensitivity curves of future GW observatories.
The data points of the projected sensitivity curves are taken from Ref.~\cite{Schmitz:2020syl}.
One can see that the spectra are deviated due to the existence of the knot dominated era (solid lines)
and can be distinguished from the conventional spectra without the knot domination (red dashed lines).

\section*{Baryogenesis}
Our Universe is baryon asymmetric, i.e., baryon is more than anti-baryon 
as characterized by a quantity $Y_B\equiv (n_B - n_{\bar{B}})/s \simeq 0.8\times 10^{-10}$~\cite{Planck:2018vyg}
where $n_{B(\bar{B})}$ and $s$ are the number density of (anti)baryon and the entropy density, respectively.
The SM however cannot explain this asymmetry.
As stated above, the knot solitons remain in the Universe and later decay into particles.
When $T_{rh}$ is larger than the electroweak (EW) scale $\mathcal{O}(10^2)$ GeV,\,\footnote{
There can be an alternative scenario of baryogenesis even when $T_{rh}$ is smaller than the electroweak scale.
This is complementary to the scenario presented here. See Appendix~\ref{app-link-baryogenesis} for the details.
}
this decay process can produce baryon asymmetry in the Universe in a similar way to non-thermal leptogenesis via inflaton decay~\cite{Asaka:1999yd,Giudice:1999fb,Asaka:1999jb,Lazarides:1990huy,Kumekawa:1994gx}.
Note that the baryon asymmetry produced at earlier stages than the end of the knot domination is diluted by the secondary reheating due to the knot decay.
For simplicity, we here assume it to be sufficiently diluted away and concentrate on the baryon asymmetry produced by the knot decay.

To see that this scenario works,
we should note that to ensure cancellation of the gauge anomaly of $U(1)_{B-L}$,
this model contains the right-handed (RH) neutrinos $\nu_R$ described by the Lagrangian
\begin{equation}
\mathcal{L}_{R} = i \bar{\nu}_{Ri} \slashed{D} \nu_{Ri}  + \left( y_R^{ij} \phi_1^\ast \nu_{Ri}^\dagger \nu^c_{Rj} + y_D^{ij} \bar\nu_{Ri} H^\dagger  \ell_j + \mathrm{h.c.}\right) \, ,
\end{equation}
where $H$ and $\ell$ are the SM Higgs doublet and lepton doublet, respectively.
 $y_R^{ij}$ ($y_D^{ij}$) is the Majorana-type (Dirac-type) Yukawa coupling constant.
The indices $i,j$ denote the generations and run as $i,j=1,2,3$.
After $\phi_1$ takes the VEV $v$, 
the RH neutrinos get Majorana masses,
whose mass eigenvalues and eigenstates are denoted by $M_{Ri}$ and $N_i$, respectively.
We suppose $M_{R1}< M_{R2}< M_{R3}$, 
so that the baryon asymmetry is dominantly produced via the lightest state $N_{1}$.
Through the couplings with $\phi_1$ and $A_\mu$ in $\mathcal{L}_{R}$, the decay of a knot soliton produces $N_i$.
We parameterize an energy fraction of the produced $N_1$ compared to the knot static energy as $f_{N_1}\in[0,1]$, 
i.e., the produced number of $N_1$ is given as $f_{N_1} M / M_{R1}$,
in which $f_{N_1}$ is not expected to be very small when $M_{R1}$ is smaller than the gauge boson mass.

Typically $N_1$ has much larger decay rate,
whose main decay channels are $N_1 \to H + \ell$ and $N_1 \to H^\dagger + \bar{\ell}$,
than that of the knot soliton, 
so that the produced $N_1$ immediately decay into $H$ ($H^\dagger$) and $\ell$ ($\bar{\ell}$).
When their decay rates are different, the lepton asymmetry $n_L - n_{\bar{L}}$ can be produced.
See Appendix~\ref{app:leptogenesis} for the calculation of the lepton asymmetry.
Through the EW sphaleron process as usual, 
some of the lepton asymmetry is converted into the baryon asymmetry.
Thus leptogenesis is followed by baryogenesis.

When the masses $M_{R1}, M_{R2}, M_{R3}$ are not degenerated,
the produced baryon asymmetry is maximized for a hierarchical case $M_{R1}\ll M_{R2}, M_{R3}$~\cite{Flanz:1994yx,Covi:1996wh,Buchmuller:1997yu,Davidson:2002qv},
leading to
\begin{align}
 Y_B 
 &\simeq  4.1 \times 10^{-11} \, f_{N_1} \left(\frac{T_{rh} }{10^6\, \mr{GeV}}\right)
 \left(\frac{m_3}{0.05 \, \mathrm{eV}}\right) \delta_\mr{eff}  \,, \label{eq:YB-non-resonant}
\end{align}
where $m_3$ and $\delta_\mr{eff}$ are the heaviest mass of the active neutrinos and the effective CP phase coming from the Yukawa matrices, respectively.
If one takes $f_{N}=\delta_\mr{eff}=\mathcal{O}(1)$, 
the correct baryon asymmetry can be explained for $T_{rh} \simeq 10^6$ GeV.
This case is however difficult to test in terms of GW spectrum by near future experiments
because it requires sensitivity for higher frequency region than $10^5$ Hz
as shown in Fig.~\ref{fig:GW}.

On the other hand, when the masses of $N_1$ and $N_2$ are almost degenerated, 
$Y_B$ can be enhanced to be as large as 
\begin{align}
 Y_B 
&\simeq 0.8 \times f_{N_1}\frac{ T_{rh}}{M_{R1}}  \label{eq:resonant}
\end{align}
like resonant leptogenesis~\cite{Pilaftsis:1997jf,Pilaftsis:2003gt}.
In such a case, the reheating temperature $T_{rh}$ can be lowered to, say, $\mathcal{O}(10^{2})$~GeV for $M_{R1}=10^{12}$~GeV.
This case is quite attractive 
because the stochastic GW spectrum lies within the range of CE, and hence this scenario is expected to be tested.
(See the orange line in Fig.~\ref{fig:GW}.)

To summarize, within reasonable setups,
the decay of the knot solitons can produce the correct baryon asymmetry,
which gives the matter-antimatter asymmetry in the present Universe.
In this sense, although the knot itself is not the building block of matter as conjectured by Kelvin, 
it might be the origin of matter and a key ingredient of nature.
Fig.~\ref{fig:history} shows the summary of cosmological history.

For further studies in the phenomenological perspective, we need concrete model building.
Particularly, it is worth considering embedding this setup into GUTs such as $SO(10)$ GUTs
since $U(1)_{B-L}$ and the SM gauge group are elegantly embedded into $SO(10)$ as is well-known.
Although we have taken $C$, the coefficient of the Chern-Simons coupling, as a free parameter,
it may be explained as one-loop effect of matter fermions charged under the $U(1)_{B-L}$ and $U(1)_\mr{PQ}$ symmetries. 
It is not obvious whether the parameter space and field contents that ensure the existence of the knot soliton is compatible with GUTs.
If so, the knot soliton can be regarded as an indirect probe of GUTs.
In any case, the knot soliton in particle physics has many potential applications to understand fundamental aspects of nature.

\section*{Acknowledgments}
The authors would like to thank
Simone Blasi, 
Asuka Ito, 
Ryusuke Jinno, 
Kohei Kamada, 
Yuka Kotorii, 
Shunya Mizoguchi,
Nobuyuki Matsumoto, 
Hiroto Shibuya, 
Fumio Uchida, 
Wen Yin and 
Ryo Yokokura for useful comments.
The numerical calculations were carried out on Yukawa-21 at YITP in Kyoto University.
This work is supported in part by JSPS Grant-in-Aid for Scientific Research KAKENHI Grant No. JP22H01221 (M.~E. and M.~N.) and No. JP22KJ3123 (Y.~H.)
and by the Deutsche Forschungsgemeinschaft under Germany's Excellence Strategy - EXC 2121 Quantum Universe - 390833306.
The work of M.~N. is supported in part by the WPI program ``Sustainability with Knotted Chiral Meta Matter (SKCM$^2$)'' at Hiroshima University.
The photo of a galaxy used in Fig.~\ref{fig:history} is taken from NASA Image and Video Library.

\section*{Appendix}

\subsection{Numerical technique to obtain knot solitons}
\label{app-numerical}

We give numerical technique performed in this work.
Because we are interested in static solutions of the EOMs,
it is sufficient to minimize the energy functional,
\begin{align}
 E & = \int d^3 x \, \mathcal{E}
\end{align}
with
\begin{align}
\mathcal{E} = &  (D_i \phi_1)^2 + (D_i \phi_2)^2 + V(\phi_1,\phi_2) + \frac{1}{2} (\partial_i A_j)^2 \nonumber  \\
& - \frac{1}{2} (\partial_i A_0)^2 - g A_0 J_0
 +  C \epsilon^{ijk} \partial_i a  A_0 \partial_j A_k  \nonumber \\
& + \frac{\gamma-1}{2} (\partial_i A_i)^2\, , \label{eq:energy-density}
\end{align}
and
\begin{equation}
J_0 \equiv g A_0(q_1^2  |\phi_1|^2 + q_2^2 |\phi_2|^2) \, ,
\end{equation}
where we have added the gauge fixing term $\gamma (\partial_i A_i)^2/2$ to realize the Coulomb gauge $\partial_i A_i = 0$
and $\gamma$ is the gauge fixing parameter.

We here meet two obstructions to perform the minimization.
\begin{itemize}
 \item The energy density~\eqref{eq:energy-density} is apparently not positive definite
because the temporal component $A_0$ has the ``wrong sign''.
 \item For large $C$, the first-order differential term $A_0 \partial_j A_k$ is dominant over the kinetic term $(\partial_i A_j)^2$.
\end{itemize}
Due to the first obstruction, a naive minimization with respect to $A_0$ does not converge.
To avoid this, we solve separately the EOM of $A_0$ called the Gauss-law constraint,
\begin{equation}
 0 = \frac{ \delta E}{\delta A_0} = \partial^2 A_0 - 2 g J_0  + C \epsilon^{ijk} \partial_i a \partial_j A_k \, ,\label{eq:Gauss-law}
\end{equation}
from which it follows that Eq.~\eqref{eq:energy-density} is rewritten as
\begin{align}
\mathcal{E} = &  (D_i \phi_1)^2 + (D_i \phi_2)^2 + V(\phi_1,\phi_2) + \frac{1}{2} (\partial_i A_j)^2 \nonumber \\
& +  \frac{1}{2}C \rho A_0 + \frac{\gamma-1}{2} (\partial_i A_i)^2  \label{eq:energy-substituted}
\end{align}
with $\rho=\epsilon^{ijk} (\partial_i a)  \partial_j A_k$.\footnote{This expression can be made obviously positive definite 
when one uses the following formal solution of the Gauss law,
\begin{equation*}
 A_0 = C \left[-\partial_i^2 + 2g^2 (q_1^2 |\phi_2|^2+q_2^2 |\phi_2|^2) \right]^{-1} \epsilon^{ijk} \partial_i a \partial_j A_k ,
\end{equation*}
from which the fifth term in Eq.~\eqref{eq:energy-substituted} becomes
\begin{align*}
 \frac{C^2}{2} \rho \, \left[-\partial_i^2 + 2g^2 (q_1^2 |\phi_2|^2+q_2^2 |\phi_2|^2) \right]^{-1} \,\rho \, .
\end{align*}
}

The second point often produces a numerical instability when we discretize the system.
To resolve this, we introduce an auxiliary field $B_i$
and replace $\rho$ as
\begin{align}
\rho \to B_i\partial_i a 
\end{align}
in Eq.~\eqref{eq:energy-substituted}
imposing constraints
\begin{equation}
B^i = \epsilon^{ijk} \partial_j A_k \quad (i=1,2,3)\, .\label{eq:constraint}
\end{equation}
Thanks to this, there is no first-order differential term apparently.
In order to realize these constraints,
we follow the procedure presented in Ref.~\cite{Hestenes1969MultiplierAG}.
We introduce multiplier fields $w_i(x)$ 
and add terms 
\begin{equation}
w_i (B^i - \epsilon^{ijk} \partial_j A_k) + \frac{U}{2} (B^i - \epsilon^{ijk} \partial_j A_k)^2
\end{equation}
to the energy density ($U=\mathrm{const.}$).
It is shown in  Ref.~\cite{Hestenes1969MultiplierAG} that 
these constraints are achieved by updating $w_i$ iteratively as follows
\begin{equation}
w_{i} \to w_i + U (B^i - \epsilon^{ijk} \partial_j A_k)\label{eq:multiplier} 
\end{equation}
when $U$ is sufficiently large.

Then the task is to minimize the three-dimensional integration of Eq.~\eqref{eq:energy-substituted}
in which $A_0$ is determined by solving the Gauss law constraint~\eqref{eq:Gauss-law}
and the constraints~\eqref{eq:constraint} are imposed.
We introduce a naive spacial discretization with lattice spacing $d$
and approximate the derivatives by second-order central-difference scheme.
Denoting the all variables on the lattice sites by $u_{a,j} = \{\mr{Re}\, \phi_{1,2}(x_j), \mr{Im}\, \phi_{1,2}(x_j),A_i(x_j),B_i(x_j)\}$ and $s_j = A_0(x_j)$,
we have $E\simeq d^3 \sum \mathcal{E}_\mr{disc}(u_{a,j},s_j)$, where the summation is taken over all lattice points.

We adopt the so-called relaxation procedure, 
in which one starts from appropriate values $u_a^{(0)}$ and
solve an iterative equation,
\begin{align}
u_{a,j}^{(n+1)} = u_{a,j}^{(n)} - \alpha \frac{\partial \mathcal{E}_\mr{disc}(u^{(n)},s^{(n)})}{\partial u_{a,j}^{(n)}}\, ,\label{eq:relaxation}
\end{align}
to obtain $(n+1)$-th step values from $n$-th step ones. 
Here, 
$\alpha$ is the step size.
To determine $s_j^{(n+1)}$, we solve the Gauss law numerically at each step by the similar procedure,
\begin{align}
  s_j^{(n+1)} & = s_j^{(n)} \nonumber \\
&\hspace{-1em}  + \beta \left[\Delta s_j^{(n)} - 2 g J_0(u_{a,j}^{(n+1)},s_j^{(n)})  + C \rho(u_{a,j}^{(n+1)},s_j^{(n)}) \right]\label{eq:Gauss-law-disc}
\end{align}
where $\Delta$ is the discretized Laplacian and $\beta$ is the step size for $s_j$.
In addition, we have to repeat the update for the multiplier $w_i$ given by Eq.~\eqref{eq:multiplier}.
To summarize, we repeat the following procedure:
\begin{equation}
 \left[\eqref{eq:relaxation} \to \eqref{eq:Gauss-law-disc} \to \eqref{eq:multiplier}\right] \to \left[\eqref{eq:relaxation} \to \eqref{eq:Gauss-law-disc} \to \eqref{eq:multiplier}\right] \to \cdots \, . \nonumber
\end{equation}
If this iteration converges,
the configuration is the static solution of the EOM with the conditions~\eqref{eq:Gauss-law} and \eqref{eq:constraint}.

In this work, we take $\gamma=1+U$, $U=50$, $d = 0.8/v$, $\alpha=4\times 10^{-4} v^{-2}$, and $\beta=2\times 10^{-3}v^{-2}$.
The number of lattice sites is $320^3$.
We use a convergence criteria that 
\begin{align}
\sum_a \overline{\left(u_{a,j}^{(n+1)} - u_{a,j}^{(n)}\right)^2} < 1.06 \times 320^{-3} \, ,
\end{align}
where the overline indicates the spatial average.

\subsection{Calculation of non-thermal leptogenesis}
\label{app:leptogenesis}
We here give more detailed calculation for leptogenesis from the knot decay.
After $N_1$'s are produced from the knots,
they further decay into $H$ ($H^\dagger$) and $\ell$ ($\bar{\ell}$).
It is convenient to introduce the so-called lepton asymmetry parameter defined through decay rates of the processes as
\begin{equation}
 \epsilon \equiv \frac{\Gamma(N_1 \to H + \ell) - \Gamma(N_1 \to H^\dagger + \bar{\ell})}{\Gamma(N_1 \to H + \ell) + \Gamma(N_1 \to H^\dagger + \bar{\ell})} \, ,
\end{equation}
which depends on parameters of the model.
Then, the produced lepton asymmetry per knot soliton is expressed as $\epsilon f_{N_1} M / M_{R1}$.

Since $T_{rh}$ is assumed to be larger than the EW scale, 
the EW sphaleron process is equilibrated 
and convert the lepton number into the baryon number as 
\begin{align}
 Y_B &= - \left.\frac{28}{79} \frac{n_L - n_{\bar{L}}}{s} \right|_{\mathrm{decay}}\nonumber \\
 &\simeq - \left.\frac{28}{79} \frac{\epsilon f_{N_1} n_\mr{knot} M}{M_{R1}} \frac{45}{2\pi}\frac{1}{g_s T_{rh}^3}\right|_{\mathrm{decay}} \nonumber\\
 &\simeq - \frac{21\pi}{79} \epsilon  f_{N_1}\frac{ T_{rh}}{M_{R1}} \, , \label{eq:YB}
\end{align}
where $g_s(\simeq g_*)$ is the effective relativistic degrees of freedom for entropy.
Note that $M_{R1}$ should satisfy $M_{R1}>T_{rh}$ to avoid the wash-out effect.

When the masses $M_{R1}, M_{R2}, M_{R3}$ are not degenerated,
the parameter $\epsilon$ is maximized for a hierarchical case $M_{R1}\ll M_{R2}, M_{R3}$,
as given by~\cite{Flanz:1994yx,Covi:1996wh,Buchmuller:1997yu,Davidson:2002qv}
\begin{equation}
 \epsilon \simeq -\frac{3}{8\pi}\frac{M_{R1}}{\langle H \rangle^2} m_3 \delta_\mr{eff}\label{eq:epsilon} \, .
\end{equation}
Substituting Eq.~\eqref{eq:epsilon} into Eq.~\eqref{eq:YB},
one gets Eq.~\eqref{eq:YB-non-resonant}.

On the other hand, when the masses of $N_1$ and $N_2$ are almost degenerated, $\epsilon$ can be enhanced to be as large as $\mathcal{O}(1)$ like resonant leptogenesis~\cite{Pilaftsis:1997jf,Pilaftsis:2003gt}.
In such a case, $Y_B$ is given by Eq.~\eqref{eq:resonant}.

\subsection{Baryogenesis via linking flux}
\label{app-link-baryogenesis}

We present an alternative scenario that the knot decay may produce baryon asymmetry in the Universe 
even when $T_{rh}$ is smaller than the EW scale $\mathcal{O}(10^2)$ GeV.
The key is the fact that 
the $\phi_2$ string also has the magnetic flux of the $U(1)_{B-L}$ gauge field and becomes a fractional vortex
when $\phi_2$ has a non-zero $B-L$ charge, $q_2 \neq 0$.
This allows the knot solitons to have the linking magnetic fluxes of the $U(1)_{B-L}$ gauge field.
This linking flux has the Chern-Simons number (or helicity) $N_{CS}[A]\sim q_1 q_2 N_\mr{link}$.

In general, this $U(1)_{B-L}$ gauge field $A_\mu$ has a kinetic mixing with the $U(1)_Y$ gauge field in the SM,
$\mathcal{L}_\mathrm{mix} = (\epsilon_\mathrm{mix}/2) F_{\mu\nu}Y^{\mu\nu}$,
especially when they arise from a unified simple gauge group such as $SO(10)$.
Due to the kinetic mixing, the magnetic flux of $A_\mu$ induces the flux of $Y_\mu$ along the $\phi_1$ and $\phi_2$ strings.
Therefore a knot soliton with the Chern-Simons number of $U(1)_X$ must contain 
the Chern-Simons number of $U(1)_Y$, $N_{CS}[Y] \sim \epsilon_\mathrm{mix}^2 N_{CS}[A] \sim \epsilon_\mathrm{mix}^2 q_1 q_2 N_\mr{link}$.
In the broken phase of the EW symmetry, the $U(1)_Y$ flux is dressed as the $Z$ boson flux~\cite{Hyde:2013fia},
which is a superposition of the $U(1)_Y$ and $SU(2)_W$ fluxes.
(The unbroken $U(1)$ electromagnetic flux is not induced.)

When the knots decay at cosmic time $t\simeq \tau_\mathrm{decay}$, all the magnetic fluxes contained in the knots disappear, 
and hence $N_{CS}[Y]$ changes from $\epsilon_\mathrm{mix}^2 q_1 q_2N_\mr{link} $ into $0$ per knot.
This decay process produces the baryon number
because the baryon number is not conserved due to the quantum anomaly
but is related with the change of the Chern-Simons number of the $U(1)_Y$ field as $\Delta B = 3 \Delta N_{CS}[Y]$.
Through this anomaly relation, 
the decay of each knot produces the baryon number $\Delta B \simeq \epsilon_\mathrm{mix}^2 q_1 q_2 N_\mr{link} $.
Therefore the linking number of the knot solitons can be a seed of the baryon number without leptogenesis.

In Refs.~\cite{Kamada:2016eeb,Kamada:2016cnb},
baryogenesis is realized via decay of helical $U(1)_Y$ magnetic fields.
While in their scenario the baryon number is produced during the EW phase transition (crossover),
we now consider the case that the baryon number is produced after completion of the EW phase transition
since $T_{rh}$ is lower than the EW scale, and hence it is not washed out by the sphaleron process.

In this scenario, we should consider carefully the creation of the knot solitons.
Indeed, when the knot solitons are produced by the Kibble-Zurek mechanism, 
anti-knot solitons, in which the $\phi_1$ and $\phi_2$ strings are linked with the relatively opposite direction compared to knot solitons,
 are necessarily produced as well.
This anti-knot has the Chern-Simons number with the opposite sign, $N_{CS}[A] \sim - \epsilon_\mathrm{mix}^2 q_1 q_2N_\mr{link} $.
If there is no probability bias between the knots and anti-knots,
their produced number should be equal, 
which results in that the net Chern-Simons number and hence the net baryon number are zero in total.
To avoid this, one needs an effective chemical potential $\mu_\mr{eff}$ between them around the phase transition $T\sim T_c$.
We assume that the asymmetry between them is given by the Boltzmann factor, leading to
\begin{align}
\left.  n_\mr{knot} - \bar n_\mr{knot} \right|_{\text{production}} 
&\sim n_\mr{knot} \frac{\mu_\mr{eff}(T_c)}{T_c}
\end{align}
within the linear perturbation assuming $\mu_\mr{eff}(T_c)/T_c \lesssim 1$.

Using this, the produced net baryon number from the decay of the (anti)knots is given as (assuming $N_\mr{link}\sim 1$)
\begin{align}
Y_B & \sim \left. \frac{n_\mr{knot} - \bar n_\mr{knot}}{s} \right|_\text{decay} \epsilon_\mathrm{mix}^2 q_1 q_2 
\nonumber \\
& \sim   \left. n_\mr{knot}\right|_\text{decay}  \,\frac{\mu_\mr{eff}(T_c)}{T_c} \, \epsilon_\mathrm{mix}^2 q_1 q_2 
\frac{45}{2\pi}\frac{1}{g_s T_{rh}^3} \nonumber \\
& \sim  10^{-10} \frac{\mu_\mr{eff}(T_c)}{T_c} \,\epsilon_\mathrm{mix}^2 q_1 q_2 \left(\frac{T_{rh}}{100 \, \mr{GeV}}\right)
\left(\frac{10^{12} \, \mathrm{GeV}}{M} \right)  \, .
\end{align}
To get the observed baryon asymmetry,
we need the mass of the knot soliton $M$ smaller than $10^{12}$ GeV.
However, $M$ is typically given as $M\sim \mathcal{O}(10^3) \,v/g$,
and $v$ should satisfy $v \gtrsim 10^{8}$ GeV,
in which the last inequality comes from astrophysical constraints on the axion decay constant, $f_a \gtrsim \mathcal{O}(10^8)$ GeV~\cite{Irastorza:2018dyq,Chang:2018rso,Hamaguchi:2018oqw,Caputo:2024oqc}.
Thus successful baryogenesis works for $10^{8} \, \mr{GeV} \lesssim v \lesssim 10^9 \, \mr{GeV}$.

We here comment on two possible origins of the chemical potential $\mu_\mr{eff}$.
The first one is a new pseudo scalar $a'$ that behaves as a time-dependent background.
If $A_\mu$ couples to $a'$ via $a' F_{\mu\nu} \tilde F^{\mu\nu}$ in the Lagrangian, 
that term becomes  $\partial_t a' \epsilon^{ijk}A_i \partial_j A_k$,
leading to $\mu_\mr{eff} \simeq \partial_t a' N_{CS}[A]$.
This resembles the Affleck-Dine mechanism~\cite{Affleck:1984fy}, Axiogenesis~\cite{Co:2019wyp},
and the production of helical magnetic fields from axion inflation~\cite{Turner:1987bw,Garretson:1992vt,Anber:2006xt}.

The second example 
of the chemical potential $\mu_\mr{eff}$
is the chiral charge asymmetry $n_5$,
which can be produced from decays of heavy particles.
For instance, the chiral charge of electrons can be produced~\cite{Kamada:2018tcs,Domcke:2020quw} in a similar way to the GUT baryogenesis~\cite{Yoshimura:1978ex,Ignatiev:1978uf,Weinberg:1979bt} and is not washed out by the electron Yukawa coupling at high temperature.
Through the chiral anomaly, this chiral asymmetry lowers (raises) energy of a helical (anti-helical) configuration of the $U(1)_Y$ gauge field,
like chiral plasma instability~\cite{Akamatsu:2013pjd}.
Thus, an imbalance between the knot and anti-knot might be induced as $\mu_\mr{eff} \simeq \mu_5 N_{CS}[Y]$ with $\mu_5 \sim n_5/T^2$.

In any case, we need some mechanism to give non-zero $\mu_\mr{eff}$ that leads to an asymmetry between knots and anti-knots.
Obviously, it is required to have a closer look at the Kibble-Zurek mechanism in the presence of $\mu_\mr{eff}$ in order to obtain precise predictions of the asymmetry.
After they are asymmetrically produced by the mechanism,
the asymmetry is stored in the (anti-)knots as the net Chern-Simons number $N_{CS}[Y]$,
and is converted into the baryon asymmetry by their decay.
In this sense, the baryon asymmetry is ``hidden'' in the knot sector until the decay, and is eventually ``released'' into the Universe.

\subsection{Calculation of gravitational wave spectrum}
We present formulae to calculate the GW spectrum from a cosmic string network.
For more details, see Refs.~\cite{Cui:2017ufi,Cui:2018rwi,Gouttenoire:2019kij}.
We assume that the network immediately approaches the scaling regime.
The strings in the network move as Brownian motion
and produce string loops by the reconnection.
We model this in terms of velocity-dependent one-scale model~\cite{Martins:1995tg,Martins:1996jp,Martins:2000cs} with a loop chopping efficiency $\bar c=0.23$.
Then, the produced number density of the loops per time at cosmic time $t_i$ is given by
\begin{equation}
 \frac{d n_\mr{loop}}{d t_i} = C_\mr{eff}(t_i) t_i^{-4},
\end{equation}
with an initial length $l=\alpha t_i$.
We take $\alpha\simeq 0.1$ and $C_\mr{eff}(t_i)=0.41$ ($5.5$) for the matter dominated (radiation dominated) Universe.

After the production, the loops oscillate and shrink by emitting GW.
The length at time $t$ is given as
\begin{equation}
 l = \alpha t_i - \Gamma G \mu(t-t_i)
\end{equation}
with the dimensionless constant $\Gamma \simeq 50$~\cite{Vilenkin:1981bx,Turok:1984cn,Quashnock:1990wv,Blanco-Pillado:2013qja,Blanco-Pillado:2017oxo}.
Each oscillation mode $k$ radiates GW with the frequency $f_\mr{emit}=2 k /l$ ($k=1,2,\cdots$).
After emission at time $\tilde t$, the frequency redshifts due to the cosmic expansion until today as
\begin{equation}
 f = \frac{a(\tilde t)}{a(t_0)}  \frac{ 2k}{l(\tilde t)} 
= \frac{a(\tilde t)}{a(t_0)}  \frac{2k}{\alpha t_i - \Gamma G \mu(\tilde t-t_i)} \, ,\label{eq:frequency}
\end{equation}
where $t_0$ is the current time.

Summing over all oscillation modes,
the energy density of GW per unit frequency observed today is given as
\begin{equation}
 \Omega_\mr{GW} (f)= \frac{f}{\rho_c} \frac{d \rho_\mr{GW}}{df} = \sum_k \Omega_\mr{GW}^{(k)} (f)\label{eq:omega}
\end{equation}
with
\begin{align}
  \Omega_\mr{GW}^{(k)} (f) &
= \frac{1}{\rho_c} \frac{2k}{f}\frac{\Gamma_k G \mu^2}{\alpha + \Gamma G \mu} \nonumber \\
& \hspace{0em} \times \int _{t_F}^{t_0} d \tilde{t} \, \frac{C_\mr{eff}(t_i)}{t_i^4} \left(\frac{a(\tilde t)}{a(t_0)}\right)^5 \left(\frac{a(t_i)}{a(\tilde t)}\right)^3 \Theta( t_i - t_F)\, ,
\end{align}
where $\rho_c = 3H_0^2/8\pi G$ is the critical density and $\Gamma_k = \Gamma k^{-4/3} / 3.6$~\cite{Blanco-Pillado:2013qja,Blanco-Pillado:2017oxo}.
$t_F$ is the formation time of the string network, which can be taken to be $0$ essentially.
We assume that the GW emission is dominated by cusps.
The relation between $\tilde t$ and $t_i$ is given by Eq.~\eqref{eq:frequency}.
$\Theta$ is the Heaviside step function.

We assume sudden transitions between the radiation dominated and knot dominated eras.
The summation over the modes $k$ is performed by the method presented in Ref.~\cite{Blasi:2020mfx}, to be taken up to $k_\mr{max}=10^{12}$,
which gives the slopes $f^{-1/3}$~\cite{Blasi:2020wpy,Blasi:2020mfx} at higher frequency regime of the spectra with the knot dominated era.
Although one may truncate the summation within smaller values,
it affects only an asymptotic behavior at high frequency.
Fig.~\ref{fig:GW} is obtained by plotting Eq.~\eqref{eq:omega} utilizing these techniques.

\bibliographystyle{jhep}
\bibliography{./references}

\providecommand{\href}[2]{#2}\begingroup\raggedright\begin{thebibliography}{100}

\bibitem{Thomson:1869}
W.~H. Thomson, \emph{{On Vortex Motion}}, {\emph{Trans. R. Soc. Edin.}
  {\bfseries 25} (1869) 217}.

\bibitem{tait1898knots}
P.~G. Tait, \emph{On knots i, ii, iii}, {\emph{Scientific papers} {\bfseries 1}
  (1898) 273}.

\bibitem{Kauffman:1991ds}
L.~H. Kauffman, \emph{{Knots and physics}}, Series on Knots \& Everything.
  World Scientific Publishing Co Pte Ltd, 1991,
  \href{https://doi.org/10.1142/4256}{10.1142/4256}.

\bibitem{Faddeev:1996zj}
L.~D. Faddeev and A.~J. Niemi, \emph{{Knots and particles}},
  \href{https://doi.org/10.1038/387058a0}{\emph{Nature} {\bfseries 387} (1997)
  58} [\href{https://arxiv.org/abs/hep-th/9610193}{{\ttfamily
  hep-th/9610193}}].

\bibitem{Moffatt_1969}
H.~K. Moffatt, \emph{The degree of knottedness of tangled vortex lines},
  \href{https://doi.org/10.1017/S0022112069000991}{\emph{Journal of Fluid
  Mechanics} {\bfseries 35} (1969) 117^^e2^^80^^93129}.

\bibitem{10.1063/1.881574}
R.~L. Ricca and M.~A. Berger, \emph{{Topological Ideas and Fluid Mechanics}},
  \href{https://doi.org/10.1063/1.881574}{\emph{Physics Today} {\bfseries 49}
  (1996) 28}.

\bibitem{Ricca:2009}
R.~L. Ricca, \emph{New developments in topological fluid mechanics},
  \href{https://doi.org/10.1393/ncc/i2009-10355-2}{\emph{Nuovo Cim.} {\bfseries
  C 32} (2009) 185}.

\bibitem{Kleckner:2013}
D.~Kleckner and W.~T.~M. Irvine, \emph{Creation and dynamics of knotted
  vortices}, \href{https://doi.org/10.1038/nphys2560}{\emph{Nature Physics}
  {\bfseries 9} (2013) 253}.

\bibitem{Arnold:2021}
V.~I. Arnold and B.~A. Khesin, \emph{Topological Methods in Hydrodynamics}.
  Springer Cham, 2021,
  \href{https://doi.org/10.1007/978-3-030-74278-2}{10.1007/978-3-030-74278-2}.

\bibitem{Babaev:2001zy}
E.~Babaev, L.~D. Faddeev and A.~J. Niemi, \emph{{Hidden symmetry and knot
  solitons in a charged two-condensate Bose system}},
  \href{https://doi.org/10.1103/PhysRevB.65.100512}{\emph{Phys. Rev. B}
  {\bfseries 65} (2002) 100512}
  [\href{https://arxiv.org/abs/cond-mat/0106152}{{\ttfamily
  cond-mat/0106152}}].

\bibitem{Rybakov:2018ktd}
F.~N. Rybakov, J.~Garaud and E.~Babaev, \emph{{Stable Hopf-Skyrme topological
  excitations in the superconducting state}},
  \href{https://doi.org/10.1103/PhysRevB.100.094515}{\emph{Phys. Rev. B}
  {\bfseries 100} (2019) 094515}
  [\href{https://arxiv.org/abs/1807.02509}{{\ttfamily 1807.02509}}].

\bibitem{PhysRevE.85.036306}
D.~Proment, M.~Onorato and C.~F. Barenghi, \emph{Vortex knots in a
  bose-einstein condensate},
  \href{https://doi.org/10.1103/PhysRevE.85.036306}{\emph{Phys. Rev. E}
  {\bfseries 85} (2012) 036306}.

\bibitem{Kleckner2016}
D.~Kleckner, L.~H. Kauffman and W.~T.~M. Irvine, \emph{How superfluid vortex
  knots untie}, \href{https://doi.org/10.1038/nphys3679}{\emph{Nature Physics}
  (2016) 650^^e2^^80^^93655}.

\bibitem{Kawaguchi:2008xi}
Y.~Kawaguchi, M.~Nitta and M.~Ueda, \emph{{Knots in a Spinor Bose-Einstein
  Condensate}},
  \href{https://doi.org/10.1103/PhysRevLett.100.180403}{\emph{Phys. Rev. Lett.}
  {\bfseries 100} (2008) 180403}
  [\href{https://arxiv.org/abs/0802.1968}{{\ttfamily 0802.1968}}].

\bibitem{Hall:2016}
D.~Hall, M.~Ray and K.~Tiurev~{\it et.al.}, \emph{Tying quantum knots},
  \href{https://doi.org/10.1038/nphys3624}{\emph{Nature Phys} {\bfseries 12}
  (2016) 478}.

\bibitem{Ollikainen:2019dyh}
T.~Ollikainen, A.~Blinova, M.~M\"ott\"onen and D.~S. Hall, \emph{{Decay of a
  Quantum Knot}},
  \href{https://doi.org/10.1103/PhysRevLett.123.163003}{\emph{Phys. Rev. Lett.}
  {\bfseries 123} (2019) 163003}
  [\href{https://arxiv.org/abs/1908.01285}{{\ttfamily 1908.01285}}].

\bibitem{Volovik:1977}
G.~Volovik and V.P.Mineev, \emph{Particle-like solitons in superfluid 3he
  phases}, {\emph{JETP} {\bfseries 46} (1977) 401}.

\bibitem{Volovik:2003fe}
G.~E. Volovik, \emph{{The Universe in a helium droplet}}, International Series
  of Monographs on Physics. Oxford Scholarship Online, 2009,
  \href{https://doi.org/10.1093/acprof:oso/9780199564842.001.0001}{10.1093/acprof:oso/9780199564842.001.0001}.

\bibitem{PhysRevLett.110.237801}
B.~G.-g. Chen, P.~J. Ackerman, G.~P. Alexander, R.~D. Kamien and I.~I.
  Smalyukh, \emph{Generating the hopf fibration experimentally in nematic
  liquid crystals},
  \href{https://doi.org/10.1103/PhysRevLett.110.237801}{\emph{Phys. Rev. Lett.}
  {\bfseries 110} (2013) 237801}.

\bibitem{PhysRevLett.113.027801}
T.~Machon and G.~P. Alexander, \emph{Knotted defects in nematic liquid
  crystals}, \href{https://doi.org/10.1103/PhysRevLett.113.027801}{\emph{Phys.
  Rev. Lett.} {\bfseries 113} (2014) 027801}.

\bibitem{Ackerman:2015}
P.~Ackerman, J.~van~de Lagemaat and I.~Smalyukh, \emph{Self-assembly and
  electrostriction of arrays and chains of hopfion particles in chiral liquid
  crystals}, \href{https://doi.org/10.1038/ncomms7012}{\emph{Nat. Comm.}
  {\bfseries 6} (2015) 6012}.

\bibitem{Ackerman:2017}
P.~Ackerman and I.~Smalyukh, \emph{Static three-dimensional topological
  solitons in fluid chiral ferromagnets and colloids},
  \href{https://doi.org/10.1038/nmat4826}{\emph{Nat. Mater.} {\bfseries 16}
  (2017) 426}.

\bibitem{Ackerman:2017b}
P.~Ackerman and I.~Smalyukh, \emph{Diversity of knot solitons in liquid
  crystals manifested by linking of preimages in torons and hopfions},
  \href{https://doi.org/10.1103/PhysRevX.7.011006}{\emph{Phys. Rev. X}
  {\bfseries 7} (2017) 011006}.

\bibitem{Tai:2018}
J.-S. Tai, P.~Ackerman and I.~Smalyukh, \emph{Topological transformations of
  hopf solitons in chiral ferromagnets and liquid crystals},
  \href{https://doi.org/10.1073/pnas.1716887115}{\emph{PNAS} {\bfseries 115}
  (2018) 921}.

\bibitem{Tai:2019}
J.-S.~B. Tai and I.~I. Smalyukh, \emph{Three-dimensional crystals of adaptive
  knots}, \href{https://doi.org/10.1126/science.aay1638}{\emph{Science}
  {\bfseries 365} (2019) 1449^^e2^^80^^931453}.

\bibitem{RevModPhys.84.497}
G.~P. Alexander, B.~G.-g. Chen, E.~A. Matsumoto and R.~D. Kamien,
  \emph{Colloquium: Disclination loops, point defects, and all that in nematic
  liquid crystals}, \href{https://doi.org/10.1103/RevModPhys.84.497}{\emph{Rev.
  Mod. Phys.} {\bfseries 84} (2012) 497}.

\bibitem{Smalyukh:2020zin}
I.~I. Smalyukh, \emph{{Review: knots and other new topological effects in
  liquid crystals and colloids}},
  \href{https://doi.org/10.1088/1361-6633/abaa39}{\emph{Rept. Prog. Phys.}
  {\bfseries 83} (2020) 106601}.

\bibitem{Smalyukh:2022}
J.-S. Wu and I.~I. Smalyukh, \emph{Hopfions, heliknotons, skyrmions, torons and
  both abelian and nonabelian vortices in chiral liquid crystals}. Taylor \&
  Francis, 2022,
  \href{https://doi.org/10.1080/21680396.2022.2040058}{10.1080/21680396.2022.2040058}.

\bibitem{Kent:2020jvm}
N.~Kent et~al., \emph{{Creation and observation of Hopfions in magnetic
  multilayer systems}},
  \href{https://doi.org/10.1038/s41467-021-21846-5}{\emph{Nature Commun.}
  {\bfseries 12} (2021) 1562}
  [\href{https://arxiv.org/abs/2010.08674}{{\ttfamily 2010.08674}}].

\bibitem{Dennis2010}
M.~R. Dennis, R.~P. King, B.~Jack, K.~O’Holleran and M.~J. Padgett,
  \emph{Isolated optical vortex knots},
  \href{https://doi.org/10.1038/nphys1504}{\emph{Nature Physics} {\bfseries 6}
  (2010) 118}.

\bibitem{Trautman:1977im}
A.~Trautman, \emph{{Solutions of the Maxwell and Yang-Mills Equations
  Associated with Hopf Fibrings}},
  \href{https://doi.org/10.1007/BF01811088}{\emph{Int. J. Theor. Phys.}
  {\bfseries 16} (1977) 561}.

\bibitem{Ranada:1989wc}
A.~F. Ranada, \emph{{A Topological Theory of the Electromagnetic Field}},
  \href{https://doi.org/10.1007/BF00401864}{\emph{Lett. Math. Phys.} {\bfseries
  18} (1989) 97}.

\bibitem{Kedia:2013bw}
H.~Kedia, I.~Bialynicki-Birula, D.~Peralta-Salas and W.~T.~M. Irvine,
  \emph{{Tying knots in light fields}},
  \href{https://doi.org/10.1103/PhysRevLett.111.150404}{\emph{Phys. Rev. Lett.}
  {\bfseries 111} (2013) 150404}
  [\href{https://arxiv.org/abs/1302.0342}{{\ttfamily 1302.0342}}].

\bibitem{Arrayas:2017sfq}
M.~Array\'as, D.~Bouwmeester and J.~L. Trueba, \emph{{Knots in
  electromagnetism}},
  \href{https://doi.org/10.1016/j.physrep.2016.11.001}{\emph{Phys. Rept.}
  {\bfseries 667} (2017) 1}.

\bibitem{Shankar2022}
S.~Shankar, A.~Souslov, M.~J. Bowick, M.~C. Marchetti and V.~Vitelli,
  \emph{Topological active matter},
  \href{https://doi.org/10.1038/s42254-022-00445-3}{\emph{Nat Rev Phys}
  {\bfseries 4} (2022) 380}.

\bibitem{Witten:1988hf}
E.~Witten, \emph{{Quantum Field Theory and the Jones Polynomial}},
  \href{https://doi.org/10.1007/BF01217730}{\emph{Commun. Math. Phys.}
  {\bfseries 121} (1989) 351}.

\bibitem{Battye:1998pe}
R.~A. Battye and P.~M. Sutcliffe, \emph{{Knots as stable soliton solutions in a
  three-dimensional classical field theory.}},
  \href{https://doi.org/10.1103/PhysRevLett.81.4798}{\emph{Phys. Rev. Lett.}
  {\bfseries 81} (1998) 4798}
  [\href{https://arxiv.org/abs/hep-th/9808129}{{\ttfamily hep-th/9808129}}].

\bibitem{Manton:2004tk}
N.~S. Manton and P.~Sutcliffe, \emph{{Topological solitons}}, Cambridge
  Monographs on Mathematical Physics. Cambridge University Press, 2004,
  \href{https://doi.org/10.1017/CBO9780511617034}{10.1017/CBO9780511617034}.

\bibitem{Radu:2008pp}
E.~Radu and M.~S. Volkov, \emph{{Existence of stationary, non-radiating ring
  solitons in field theory: knots and vortons}},
  \href{https://doi.org/10.1016/j.physrep.2008.07.002}{\emph{Phys. Rept.}
  {\bfseries 468} (2008) 101}
  [\href{https://arxiv.org/abs/0804.1357}{{\ttfamily 0804.1357}}].

\bibitem{Kobayashi:2013xoa}
M.~Kobayashi and M.~Nitta, \emph{{Torus knots as Hopfions}},
  \href{https://doi.org/10.1016/j.physletb.2013.12.002}{\emph{Phys. Lett. B}
  {\bfseries 728} (2014) 314}
  [\href{https://arxiv.org/abs/1304.6021}{{\ttfamily 1304.6021}}].

\bibitem{Shnir:2018yzp}
Y.~M. Shnir, \emph{{Topological and Non-Topological Solitons in Scalar Field
  Theories}}. Cambridge University Press, 7, 2018,
  \href{https://doi.org/10.1017/9781108555623}{10.1017/9781108555623}.

\bibitem{Babaev:2004rm}
E.~Babaev, A.~Sudbo and N.~W. Ashcroft, \emph{{A Superconductor to superfluid
  phase transition in liquid metallic hydrogen}},
  \href{https://doi.org/10.1038/nature02910}{\emph{Nature} {\bfseries 431}
  (2004) 666} [\href{https://arxiv.org/abs/cond-mat/0410408}{{\ttfamily
  cond-mat/0410408}}].

\bibitem{Smiseth:2004na}
J.~Smiseth, E.~Smorgrav, E.~Babaev and A.~Sudbo, \emph{{Field and temperature
  induced topological phase transitions in the three-dimensional N-component
  London superconductor}},
  \href{https://doi.org/10.1103/PhysRevB.71.214509}{\emph{Phys. Rev. B}
  {\bfseries 71} (2005) 214509}
  [\href{https://arxiv.org/abs/cond-mat/0411761}{{\ttfamily
  cond-mat/0411761}}].

\bibitem{Babaev:2002wa}
E.~Babaev, \emph{{Andreev-Bashkin effect and knot solitons in neutron stars}},
  \href{https://doi.org/10.1103/PhysRevD.70.043001}{\emph{Phys. Rev. D}
  {\bfseries 70} (2004) 043001}
  [\href{https://arxiv.org/abs/astro-ph/0211345}{{\ttfamily
  astro-ph/0211345}}].

\bibitem{Peccei:1977hh}
R.~Peccei and H.~R. Quinn, \emph{{CP Conservation in the Presence of
  Instantons}}, \href{https://doi.org/10.1103/PhysRevLett.38.1440}{\emph{Phys.
  Rev. Lett.} {\bfseries 38} (1977) 1440}.

\bibitem{Peccei:1977ur}
R.~Peccei and H.~R. Quinn, \emph{{Constraints Imposed by CP Conservation in the
  Presence of Instantons}},
  \href{https://doi.org/10.1103/PhysRevD.16.1791}{\emph{Phys. Rev. D}
  {\bfseries 16} (1977) 1791}.

\bibitem{Hidaka:2021mml}
Y.~Hidaka, M.~Nitta and R.~Yokokura, \emph{{Topological axion electrodynamics
  and 4-group symmetry}},
  \href{https://doi.org/10.1016/j.physletb.2021.136762}{\emph{Phys. Lett. B}
  {\bfseries 823} (2021) 136762}
  [\href{https://arxiv.org/abs/2107.08753}{{\ttfamily 2107.08753}}].

\bibitem{Hidaka:2021kkf}
Y.~Hidaka, M.~Nitta and R.~Yokokura, \emph{{Global 4-group symmetry and
  \textquoteright{}t Hooft anomalies in topological axion electrodynamics}},
  \href{https://doi.org/10.1093/ptep/ptab150}{\emph{PTEP} {\bfseries 2022}
  (2022) 04A109} [\href{https://arxiv.org/abs/2108.12564}{{\ttfamily
  2108.12564}}].

\bibitem{Hidaka:2020izy}
Y.~Hidaka, M.~Nitta and R.~Yokokura, \emph{{Global 3-group symmetry and 't
  Hooft anomalies in axion electrodynamics}},
  \href{https://doi.org/10.1007/JHEP01(2021)173}{\emph{JHEP} {\bfseries 01}
  (2021) 173} [\href{https://arxiv.org/abs/2009.14368}{{\ttfamily
  2009.14368}}].

\bibitem{Hidaka:2020iaz}
Y.~Hidaka, M.~Nitta and R.~Yokokura, \emph{{Higher-form symmetries and 3-group
  in axion electrodynamics}},
  \href{https://doi.org/10.1016/j.physletb.2020.135672}{\emph{Phys. Lett. B}
  {\bfseries 808} (2020) 135672}
  [\href{https://arxiv.org/abs/2006.12532}{{\ttfamily 2006.12532}}].

\bibitem{Wilczek:1987mv}
F.~Wilczek, \emph{{Two Applications of Axion Electrodynamics}},
  \href{https://doi.org/10.1103/PhysRevLett.58.1799}{\emph{Phys. Rev. Lett.}
  {\bfseries 58} (1987) 1799}.

\bibitem{Qi:2012cs}
X.-L. Qi, E.~Witten and S.-C. Zhang, \emph{{Axion topological field theory of
  topological superconductors}},
  \href{https://doi.org/10.1103/PhysRevB.87.134519}{\emph{Phys. Rev. B}
  {\bfseries 87} (2013) 134519}
  [\href{https://arxiv.org/abs/1206.1407}{{\ttfamily 1206.1407}}].

\bibitem{Stone:2016pof}
M.~Stone and P.~L.~S. Lopes, \emph{{Effective action and electromagnetic
  response of topological superconductors and Majorana-mass Weyl fermions}},
  \href{https://doi.org/10.1103/PhysRevB.93.174501}{\emph{Phys. Rev. B}
  {\bfseries 93} (2016) 174501}
  [\href{https://arxiv.org/abs/1601.07869}{{\ttfamily 1601.07869}}].

\bibitem{Stalhammar:2021tcq}
M.~St\r{a}lhammar, M.~Stone, M.~Sato and T.~H. Hansson, \emph{{Electromagnetic
  response of topological superconductors}},
  \href{https://doi.org/10.1103/PhysRevB.103.235427}{\emph{Phys. Rev. B}
  {\bfseries 103} (2021) 235427}
  [\href{https://arxiv.org/abs/2103.08960}{{\ttfamily 2103.08960}}].

\bibitem{Horvathy:2007ps}
P.~A. Horvathy, \emph{{Lectures on (abelian) Chern-Simons vortices}},  4, 2007,
  \href{https://arxiv.org/abs/0704.3220}{{\ttfamily 0704.3220}}.

\bibitem{Horvathy:2008hd}
P.~A. Horvathy and P.~Zhang, \emph{{Vortices in (abelian) Chern-Simons gauge
  theory}}, \href{https://doi.org/10.1016/j.physrep.2009.07.003}{\emph{Phys.
  Rept.} {\bfseries 481} (2009) 83}
  [\href{https://arxiv.org/abs/0811.2094}{{\ttfamily 0811.2094}}].

\bibitem{Yokoi_2013}
N.~Yokoi, \emph{Cross helicity and related dynamo},
  \href{https://doi.org/10.1080/03091929.2012.754022}{\emph{Geophysical ans
  Astrophysical Fluid Dynamics} {\bfseries 107} (2013) 114^^e2^^80^^93184}.

\bibitem{Nastase:2022aps}
H.~Nastase and J.~Sonnenschein, \emph{{Fluid-electromagnetic helicities and
  knotted solutions of the fluid-electromagnetic equations}},
  \href{https://doi.org/10.1007/JHEP12(2022)144}{\emph{JHEP} {\bfseries 12}
  (2022) 144} [\href{https://arxiv.org/abs/2205.10415}{{\ttfamily
  2205.10415}}].

\bibitem{Davis:1988ij}
R.~L. Davis and E.~P.~S. Shellard, \emph{{Cosmic Vortons}},
  \href{https://doi.org/10.1016/0550-3213(89)90594-4}{\emph{Nucl. Phys. B}
  {\bfseries 323} (1989) 209}.

\bibitem{Davis:1988jq}
R.~L. Davis and E.~P.~S. Shellard, \emph{{The Physics of Vortex
  Superconductivity. 2}},
  \href{https://doi.org/10.1016/0370-2693(88)91178-1}{\emph{Phys. Lett. B}
  {\bfseries 209} (1988) 485}.

\bibitem{Battye:2008mm}
R.~A. Battye and P.~M. Sutcliffe, \emph{{Vorton construction and dynamics}},
  \href{https://doi.org/10.1016/j.nuclphysb.2009.01.021}{\emph{Nucl. Phys. B}
  {\bfseries 814} (2009) 180}
  [\href{https://arxiv.org/abs/0812.3239}{{\ttfamily 0812.3239}}].

\bibitem{Garaud:2013iba}
J.~Garaud, E.~Radu and M.~S. Volkov, \emph{{Stable Cosmic Vortons}},
  \href{https://doi.org/10.1103/PhysRevLett.111.171602}{\emph{Phys. Rev. Lett.}
  {\bfseries 111} (2013) 171602}
  [\href{https://arxiv.org/abs/1303.3044}{{\ttfamily 1303.3044}}].

\bibitem{Battye:2021sji}
R.~A. Battye and S.~J. Cotterill, \emph{{Stable Cosmic Vortons in Bosonic Field
  Theory}}, \href{https://doi.org/10.1103/PhysRevLett.127.241601}{\emph{Phys.
  Rev. Lett.} {\bfseries 127} (2021) 241601}
  [\href{https://arxiv.org/abs/2111.07822}{{\ttfamily 2111.07822}}].

\bibitem{Battye:2021kbd}
R.~A. Battye, S.~J. Cotterill and J.~A. Pearson, \emph{{A detailed study of the
  stability of vortons}},
  \href{https://doi.org/10.1007/JHEP04(2022)005}{\emph{JHEP} {\bfseries 04}
  (2022) 005} [\href{https://arxiv.org/abs/2112.08066}{{\ttfamily
  2112.08066}}].

\bibitem{Ibe:2021ctf}
M.~Ibe, S.~Kobayashi, Y.~Nakayama and S.~Shirai, \emph{{On Stability of
  Fermionic Superconducting Current in Cosmic String}},
  \href{https://doi.org/10.1007/JHEP05(2021)217}{\emph{JHEP} {\bfseries 05}
  (2021) 217} [\href{https://arxiv.org/abs/2102.05412}{{\ttfamily
  2102.05412}}].

\bibitem{Abe:2022rrh}
Y.~Abe, Y.~Hamada, K.~Saji and K.~Yoshioka, \emph{{Quantum current dissipation
  in superconducting strings and vortons}},
  \href{https://doi.org/10.1007/JHEP02(2023)004}{\emph{JHEP} {\bfseries 02}
  (2023) 004} [\href{https://arxiv.org/abs/2209.03223}{{\ttfamily
  2209.03223}}].

\bibitem{Gudnason:2020luj}
S.~B. Gudnason and M.~Nitta, \emph{{Linking number of vortices as baryon
  number}}, \href{https://doi.org/10.1103/PhysRevD.101.065011}{\emph{Phys. Rev.
  D} {\bfseries 101} (2020) 065011}
  [\href{https://arxiv.org/abs/2002.01762}{{\ttfamily 2002.01762}}].

\bibitem{Gudnason:2020qkd}
S.~B. Gudnason and M.~Nitta, \emph{{Linked vortices as baryons in the miscible
  BEC-Skyrme model}},
  \href{https://doi.org/10.1103/PhysRevD.102.045022}{\emph{Phys. Rev. D}
  {\bfseries 102} (2020) 045022}
  [\href{https://arxiv.org/abs/2006.04067}{{\ttfamily 2006.04067}}].

\bibitem{Ruostekoski:2001fc}
J.~Ruostekoski and J.~R. Anglin, \emph{{Creating vortex rings and
  three-dimensional skyrmions in Bose-Einstein condensates}},
  \href{https://doi.org/10.1103/PhysRevLett.86.3934}{\emph{Phys. Rev. Lett.}
  {\bfseries 86} (2001) 3934}
  [\href{https://arxiv.org/abs/cond-mat/0103310}{{\ttfamily
  cond-mat/0103310}}].

\bibitem{Battye:2001ec}
R.~A. Battye, N.~R. Cooper and P.~M. Sutcliffe, \emph{{Stable skyrmions in two
  component Bose-Einstein condensates}},
  \href{https://doi.org/10.1103/PhysRevLett.88.080401}{\emph{Phys. Rev. Lett.}
  {\bfseries 88} (2002) 080401}
  [\href{https://arxiv.org/abs/cond-mat/0109448}{{\ttfamily
  cond-mat/0109448}}].

\bibitem{Nitta:2012hy}
M.~Nitta, K.~Kasamatsu, M.~Tsubota and H.~Takeuchi, \emph{{Creating vortons and
  three-dimensional skyrmions from domain wall annihilation with stretched
  vortices in Bose-Einstein condensates}},
  \href{https://doi.org/10.1103/PhysRevA.85.053639}{\emph{Phys. Rev. A}
  {\bfseries 85} (2012) 053639}
  [\href{https://arxiv.org/abs/1203.4896}{{\ttfamily 1203.4896}}].

\bibitem{Eto:2011wp}
M.~Eto, K.~Kasamatsu, M.~Nitta, H.~Takeuchi and M.~Tsubota, \emph{{Interaction
  of half-quantized vortices in two-component Bose-Einstein condensates}},
  \href{https://doi.org/10.1103/PhysRevA.83.063603}{\emph{Phys. Rev. A}
  {\bfseries 83} (2011) 063603}
  [\href{https://arxiv.org/abs/1103.6144}{{\ttfamily 1103.6144}}].

\bibitem{Ooguri:2011aa}
H.~Ooguri and M.~Oshikawa, \emph{{Instability in magnetic materials with
  dynamical axion field}},
  \href{https://doi.org/10.1103/PhysRevLett.108.161803}{\emph{Phys. Rev. Lett.}
  {\bfseries 108} (2012) 161803}
  [\href{https://arxiv.org/abs/1112.1414}{{\ttfamily 1112.1414}}].

\bibitem{ParticleDataGroup:2022pth}
{\scshape Particle Data Group} collaboration, \emph{{Review of Particle
  Physics}}, \href{https://doi.org/10.1093/ptep/ptac097}{\emph{PTEP} {\bfseries
  2022} (2022) 083C01}.

\bibitem{Weinberg:1977ma}
S.~Weinberg, \emph{{A New Light Boson?}},
  \href{https://doi.org/10.1103/PhysRevLett.40.223}{\emph{Phys. Rev. Lett.}
  {\bfseries 40} (1978) 223}.

\bibitem{Wilczek:1977pj}
F.~Wilczek, \emph{{Problem of Strong $P$ and $T$ Invariance in the Presence of
  Instantons}}, \href{https://doi.org/10.1103/PhysRevLett.40.279}{\emph{Phys.
  Rev. Lett.} {\bfseries 40} (1978) 279}.

\bibitem{Minkowski:1977sc}
P.~Minkowski, \emph{{$\mu \to e\gamma$ at a Rate of One Out of $10^{9}$ Muon
  Decays?}}, \href{https://doi.org/10.1016/0370-2693(77)90435-X}{\emph{Phys.
  Lett. B} {\bfseries 67} (1977) 421}.

\bibitem{Gell-Mann:1979vob}
M.~Gell-Mann, P.~Ramond and R.~Slansky, \emph{{Complex Spinors and Unified
  Theories}}, {\emph{Conf. Proc. C} {\bfseries 790927} (1979) 315}
  [\href{https://arxiv.org/abs/1306.4669}{{\ttfamily 1306.4669}}].

\bibitem{Mohapatra:1979ia}
R.~N. Mohapatra and G.~Senjanovic, \emph{{Neutrino Mass and Spontaneous Parity
  Nonconservation}},
  \href{https://doi.org/10.1103/PhysRevLett.44.912}{\emph{Phys. Rev. Lett.}
  {\bfseries 44} (1980) 912}.

\bibitem{Yanagida:1979as}
T.~Yanagida, \emph{{Horizontal gauge symmetry and masses of neutrinos}},
  {\emph{Conf. Proc. C} {\bfseries 7902131} (1979) 95}.

\bibitem{Fukuda:2017ylt}
H.~Fukuda, M.~Ibe, M.~Suzuki and T.~T. Yanagida, \emph{{A ''gauged'' $U(1)$
  Peccei\textendash{}Quinn symmetry}},
  \href{https://doi.org/10.1016/j.physletb.2017.05.071}{\emph{Phys. Lett. B}
  {\bfseries 771} (2017) 327}
  [\href{https://arxiv.org/abs/1703.01112}{{\ttfamily 1703.01112}}].

\bibitem{Ibe:2018hir}
M.~Ibe, M.~Suzuki and T.~T. Yanagida, \emph{{$B-L$ as a Gauged Peccei-Quinn
  Symmetry}}, \href{https://doi.org/10.1007/JHEP08(2018)049}{\emph{JHEP}
  {\bfseries 08} (2018) 049}
  [\href{https://arxiv.org/abs/1805.10029}{{\ttfamily 1805.10029}}].

\bibitem{Kim:1979if}
J.~E. Kim, \emph{{Weak Interaction Singlet and Strong CP Invariance}},
  \href{https://doi.org/10.1103/PhysRevLett.43.103}{\emph{Phys. Rev. Lett.}
  {\bfseries 43} (1979) 103}.

\bibitem{Shifman:1979if}
M.~A. Shifman, A.~Vainshtein and V.~I. Zakharov, \emph{{Can Confinement Ensure
  Natural CP Invariance of Strong Interactions?}},
  \href{https://doi.org/10.1016/0550-3213(80)90209-6}{\emph{Nucl. Phys. B}
  {\bfseries 166} (1980) 493}.

\bibitem{Irastorza:2018dyq}
I.~G. Irastorza and J.~Redondo, \emph{{New experimental approaches in the
  search for axion-like particles}},
  \href{https://doi.org/10.1016/j.ppnp.2018.05.003}{\emph{Prog. Part. Nucl.
  Phys.} {\bfseries 102} (2018) 89}
  [\href{https://arxiv.org/abs/1801.08127}{{\ttfamily 1801.08127}}].

\bibitem{Chang:2018rso}
J.~H. Chang, R.~Essig and S.~D. McDermott, \emph{{Supernova 1987A Constraints
  on Sub-GeV Dark Sectors, Millicharged Particles, the QCD Axion, and an
  Axion-like Particle}},
  \href{https://doi.org/10.1007/JHEP09(2018)051}{\emph{JHEP} {\bfseries 09}
  (2018) 051} [\href{https://arxiv.org/abs/1803.00993}{{\ttfamily
  1803.00993}}].

\bibitem{Hamaguchi:2018oqw}
K.~Hamaguchi, N.~Nagata, K.~Yanagi and J.~Zheng, \emph{{Limit on the Axion
  Decay Constant from the Cooling Neutron Star in Cassiopeia A}},
  \href{https://doi.org/10.1103/PhysRevD.98.103015}{\emph{Phys. Rev. D}
  {\bfseries 98} (2018) 103015}
  [\href{https://arxiv.org/abs/1806.07151}{{\ttfamily 1806.07151}}].

\bibitem{Caputo:2024oqc}
A.~Caputo and G.~Raffelt, \emph{{Astrophysical Axion Bounds: The 2024
  Edition}}, \href{https://doi.org/10.22323/1.454.0041}{\emph{PoS} {\bfseries
  COSMICWISPers} (2024) 041}
  [\href{https://arxiv.org/abs/2401.13728}{{\ttfamily 2401.13728}}].

\bibitem{Kibble:1980mv}
T.~W.~B. Kibble, \emph{{Some Implications of a Cosmological Phase Transition}},
  \href{https://doi.org/10.1016/0370-1573(80)90091-5}{\emph{Phys. Rept.}
  {\bfseries 67} (1980) 183}.

\bibitem{Zurek:1985qw}
W.~H. Zurek, \emph{{Cosmological Experiments in Superfluid Helium?}},
  \href{https://doi.org/10.1038/317505a0}{\emph{Nature} {\bfseries 317} (1985)
  505}.

\bibitem{Turok:1989ai}
N.~Turok, \emph{{Global Texture as the Origin of Cosmic Structure}},
  \href{https://doi.org/10.1103/PhysRevLett.63.2625}{\emph{Phys. Rev. Lett.}
  {\bfseries 63} (1989) 2625}.

\bibitem{Borrill:1991mv}
J.~Borrill, E.~J. Copeland and A.~R. Liddle, \emph{{Initial conditions for
  global texture}},
  \href{https://doi.org/10.1016/0370-2693(91)91091-9}{\emph{Phys. Lett. B}
  {\bfseries 258} (1991) 310}.

\bibitem{Leese:1991gt}
R.~A. Leese and T.~Prokopec, \emph{{Monte Carlo simulation of texture
  formation}}, \href{https://doi.org/10.1103/PhysRevD.44.3749}{\emph{Phys. Rev.
  D} {\bfseries 44} (1991) 3749}.

\bibitem{Vilenkin:2000jqa}
A.~Vilenkin and E.~S. Shellard, \emph{{Cosmic Strings and Other Topological
  Defects}}. Cambridge University Press, 7, 2000.

\bibitem{Vachaspati:1994ng}
T.~Vachaspati and G.~B. Field, \emph{{Electroweak string configurations with
  baryon number}},
  \href{https://doi.org/10.1103/PhysRevLett.73.373}{\emph{Phys. Rev. Lett.}
  {\bfseries 73} (1994) 373}
  [\href{https://arxiv.org/abs/hep-ph/9401220}{{\ttfamily hep-ph/9401220}}].

\bibitem{Kolb:1990vq}
E.~W. Kolb and M.~S. Turner, \emph{{The Early Universe}}, vol.~69. 1990,
  \href{https://doi.org/10.1201/9780429492860}{10.1201/9780429492860}.

\bibitem{Kawasaki:1999na}
M.~Kawasaki, K.~Kohri and N.~Sugiyama, \emph{{Cosmological constraints on late
  time entropy production}},
  \href{https://doi.org/10.1103/PhysRevLett.82.4168}{\emph{Phys. Rev. Lett.}
  {\bfseries 82} (1999) 4168}
  [\href{https://arxiv.org/abs/astro-ph/9811437}{{\ttfamily
  astro-ph/9811437}}].

\bibitem{Kawasaki:2000en}
M.~Kawasaki, K.~Kohri and N.~Sugiyama, \emph{{MeV scale reheating temperature
  and thermalization of neutrino background}},
  \href{https://doi.org/10.1103/PhysRevD.62.023506}{\emph{Phys. Rev. D}
  {\bfseries 62} (2000) 023506}
  [\href{https://arxiv.org/abs/astro-ph/0002127}{{\ttfamily
  astro-ph/0002127}}].

\bibitem{Hannestad:2004px}
S.~Hannestad, \emph{{What is the lowest possible reheating temperature?}},
  \href{https://doi.org/10.1103/PhysRevD.70.043506}{\emph{Phys. Rev. D}
  {\bfseries 70} (2004) 043506}
  [\href{https://arxiv.org/abs/astro-ph/0403291}{{\ttfamily
  astro-ph/0403291}}].

\bibitem{Cui:2017ufi}
Y.~Cui, M.~Lewicki, D.~E. Morrissey and J.~D. Wells, \emph{{Cosmic Archaeology
  with Gravitational Waves from Cosmic Strings}},
  \href{https://doi.org/10.1103/PhysRevD.97.123505}{\emph{Phys. Rev. D}
  {\bfseries 97} (2018) 123505}
  [\href{https://arxiv.org/abs/1711.03104}{{\ttfamily 1711.03104}}].

\bibitem{Cui:2018rwi}
Y.~Cui, M.~Lewicki, D.~E. Morrissey and J.~D. Wells, \emph{{Probing the pre-BBN
  universe with gravitational waves from cosmic strings}},
  \href{https://doi.org/10.1007/JHEP01(2019)081}{\emph{JHEP} {\bfseries 01}
  (2019) 081} [\href{https://arxiv.org/abs/1808.08968}{{\ttfamily
  1808.08968}}].

\bibitem{Gouttenoire:2019kij}
Y.~Gouttenoire, G.~Servant and P.~Simakachorn, \emph{{Beyond the Standard
  Models with Cosmic Strings}},
  \href{https://doi.org/10.1088/1475-7516/2020/07/032}{\emph{JCAP} {\bfseries
  07} (2020) 032} [\href{https://arxiv.org/abs/1912.02569}{{\ttfamily
  1912.02569}}].

\bibitem{NANOGrav:2023gor}
{\scshape NANOGrav} collaboration, \emph{{The NANOGrav 15 yr Data Set: Evidence
  for a Gravitational-wave Background}},
  \href{https://doi.org/10.3847/2041-8213/acdac6}{\emph{Astrophys. J. Lett.}
  {\bfseries 951} (2023) L8}
  [\href{https://arxiv.org/abs/2306.16213}{{\ttfamily 2306.16213}}].

\bibitem{NANOGrav:2023hvm}
{\scshape NANOGrav} collaboration, \emph{{The NANOGrav 15 yr Data Set: Search
  for Signals from New Physics}},
  \href{https://doi.org/10.3847/2041-8213/acdc91}{\emph{Astrophys. J. Lett.}
  {\bfseries 951} (2023) L11}
  [\href{https://arxiv.org/abs/2306.16219}{{\ttfamily 2306.16219}}].

\bibitem{Schmitz:2020syl}
K.~Schmitz, \emph{{New Sensitivity Curves for Gravitational-Wave Signals from
  Cosmological Phase Transitions}},
  \href{https://doi.org/10.1007/JHEP01(2021)097}{\emph{JHEP} {\bfseries 01}
  (2021) 097} [\href{https://arxiv.org/abs/2002.04615}{{\ttfamily
  2002.04615}}].

\bibitem{Kawamura:2020pcg}
S.~Kawamura et~al., \emph{{Current status of space gravitational wave antenna
  DECIGO and B-DECIGO}},
  \href{https://doi.org/10.1093/ptep/ptab019}{\emph{PTEP} {\bfseries 2021}
  (2021) 05A105} [\href{https://arxiv.org/abs/2006.13545}{{\ttfamily
  2006.13545}}].

\bibitem{Reitze:2019iox}
D.~Reitze et~al., \emph{{Cosmic Explorer: The U.S. Contribution to
  Gravitational-Wave Astronomy beyond LIGO}}, {\emph{Bull. Am. Astron. Soc.}
  {\bfseries 51} (2019) 035}
  [\href{https://arxiv.org/abs/1907.04833}{{\ttfamily 1907.04833}}].

\bibitem{Bartolo:2016ami}
N.~Bartolo et~al., \emph{{Science with the space-based interferometer LISA. IV:
  Probing inflation with gravitational waves}},
  \href{https://doi.org/10.1088/1475-7516/2016/12/026}{\emph{JCAP} {\bfseries
  12} (2016) 026} [\href{https://arxiv.org/abs/1610.06481}{{\ttfamily
  1610.06481}}].

\bibitem{Janssen:2014dka}
G.~Janssen et~al., \emph{{Gravitational wave astronomy with the SKA}},
  \href{https://doi.org/10.22323/1.215.0037}{\emph{PoS} {\bfseries AASKA14}
  (2015) 037} [\href{https://arxiv.org/abs/1501.00127}{{\ttfamily
  1501.00127}}].

\bibitem{Planck:2018vyg}
{\scshape Planck} collaboration, \emph{{Planck 2018 results. VI. Cosmological
  parameters}},
  \href{https://doi.org/10.1051/0004-6361/201833910}{\emph{Astron. Astrophys.}
  {\bfseries 641} (2020) A6}
  [\href{https://arxiv.org/abs/1807.06209}{{\ttfamily 1807.06209}}].

\bibitem{Asaka:1999yd}
T.~Asaka, K.~Hamaguchi, M.~Kawasaki and T.~Yanagida, \emph{{Leptogenesis in
  inflaton decay}},
  \href{https://doi.org/10.1016/S0370-2693(99)01020-5}{\emph{Phys. Lett. B}
  {\bfseries 464} (1999) 12}
  [\href{https://arxiv.org/abs/hep-ph/9906366}{{\ttfamily hep-ph/9906366}}].

\bibitem{Giudice:1999fb}
G.~F. Giudice, M.~Peloso, A.~Riotto and I.~Tkachev, \emph{{Production of
  massive fermions at preheating and leptogenesis}},
  \href{https://doi.org/10.1088/1126-6708/1999/08/014}{\emph{JHEP} {\bfseries
  08} (1999) 014} [\href{https://arxiv.org/abs/hep-ph/9905242}{{\ttfamily
  hep-ph/9905242}}].

\bibitem{Asaka:1999jb}
T.~Asaka, K.~Hamaguchi, M.~Kawasaki and T.~Yanagida, \emph{{Leptogenesis in
  inflationary universe}},
  \href{https://doi.org/10.1103/PhysRevD.61.083512}{\emph{Phys. Rev. D}
  {\bfseries 61} (2000) 083512}
  [\href{https://arxiv.org/abs/hep-ph/9907559}{{\ttfamily hep-ph/9907559}}].

\bibitem{Lazarides:1990huy}
G.~Lazarides and Q.~Shafi, \emph{{Origin of matter in the inflationary
  cosmology}}, \href{https://doi.org/10.1016/0370-2693(91)91090-I}{\emph{Phys.
  Lett. B} {\bfseries 258} (1991) 305}.

\bibitem{Kumekawa:1994gx}
K.~Kumekawa, T.~Moroi and T.~Yanagida, \emph{{Flat potential for inflaton with
  a discrete R invariance in supergravity}},
  \href{https://doi.org/10.1143/PTP.92.437}{\emph{Prog. Theor. Phys.}
  {\bfseries 92} (1994) 437}
  [\href{https://arxiv.org/abs/hep-ph/9405337}{{\ttfamily hep-ph/9405337}}].

\bibitem{Flanz:1994yx}
M.~Flanz, E.~A. Paschos and U.~Sarkar, \emph{{Baryogenesis from a lepton
  asymmetric universe}},
  \href{https://doi.org/10.1016/0370-2693(94)01555-Q}{\emph{Phys. Lett. B}
  {\bfseries 345} (1995) 248}
  [\href{https://arxiv.org/abs/hep-ph/9411366}{{\ttfamily hep-ph/9411366}}].

\bibitem{Covi:1996wh}
L.~Covi, E.~Roulet and F.~Vissani, \emph{{CP violating decays in leptogenesis
  scenarios}}, \href{https://doi.org/10.1016/0370-2693(96)00817-9}{\emph{Phys.
  Lett. B} {\bfseries 384} (1996) 169}
  [\href{https://arxiv.org/abs/hep-ph/9605319}{{\ttfamily hep-ph/9605319}}].

\bibitem{Buchmuller:1997yu}
W.~Buchmuller and M.~Plumacher, \emph{{CP asymmetry in Majorana neutrino
  decays}}, \href{https://doi.org/10.1016/S0370-2693(97)01548-7}{\emph{Phys.
  Lett. B} {\bfseries 431} (1998) 354}
  [\href{https://arxiv.org/abs/hep-ph/9710460}{{\ttfamily hep-ph/9710460}}].

\bibitem{Davidson:2002qv}
S.~Davidson and A.~Ibarra, \emph{{A Lower bound on the right-handed neutrino
  mass from leptogenesis}},
  \href{https://doi.org/10.1016/S0370-2693(02)01735-5}{\emph{Phys. Lett. B}
  {\bfseries 535} (2002) 25}
  [\href{https://arxiv.org/abs/hep-ph/0202239}{{\ttfamily hep-ph/0202239}}].

\bibitem{Pilaftsis:1997jf}
A.~Pilaftsis, \emph{{CP violation and baryogenesis due to heavy Majorana
  neutrinos}}, \href{https://doi.org/10.1103/PhysRevD.56.5431}{\emph{Phys. Rev.
  D} {\bfseries 56} (1997) 5431}
  [\href{https://arxiv.org/abs/hep-ph/9707235}{{\ttfamily hep-ph/9707235}}].

\bibitem{Pilaftsis:2003gt}
A.~Pilaftsis and T.~E.~J. Underwood, \emph{{Resonant leptogenesis}},
  \href{https://doi.org/10.1016/j.nuclphysb.2004.05.029}{\emph{Nucl. Phys. B}
  {\bfseries 692} (2004) 303}
  [\href{https://arxiv.org/abs/hep-ph/0309342}{{\ttfamily hep-ph/0309342}}].

\bibitem{Hestenes1969MultiplierAG}
M.~R. Hestenes, \emph{Multiplier and gradient methods}, {\emph{Journal of
  Optimization Theory and Applications} {\bfseries 4} (1969) 303}.

\bibitem{Hyde:2013fia}
J.~M. Hyde, A.~J. Long and T.~Vachaspati, \emph{{Dark Strings and their
  Couplings to the Standard Model}},
  \href{https://doi.org/10.1103/PhysRevD.89.065031}{\emph{Phys. Rev. D}
  {\bfseries 89} (2014) 065031}
  [\href{https://arxiv.org/abs/1312.4573}{{\ttfamily 1312.4573}}].

\bibitem{Kamada:2016eeb}
K.~Kamada and A.~J. Long, \emph{{Baryogenesis from decaying magnetic
  helicity}}, \href{https://doi.org/10.1103/PhysRevD.94.063501}{\emph{Phys.
  Rev. D} {\bfseries 94} (2016) 063501}
  [\href{https://arxiv.org/abs/1606.08891}{{\ttfamily 1606.08891}}].

\bibitem{Kamada:2016cnb}
K.~Kamada and A.~J. Long, \emph{{Evolution of the Baryon Asymmetry through the
  Electroweak Crossover in the Presence of a Helical Magnetic Field}},
  \href{https://doi.org/10.1103/PhysRevD.94.123509}{\emph{Phys. Rev. D}
  {\bfseries 94} (2016) 123509}
  [\href{https://arxiv.org/abs/1610.03074}{{\ttfamily 1610.03074}}].

\bibitem{Affleck:1984fy}
I.~Affleck and M.~Dine, \emph{{A New Mechanism for Baryogenesis}},
  \href{https://doi.org/10.1016/0550-3213(85)90021-5}{\emph{Nucl. Phys. B}
  {\bfseries 249} (1985) 361}.

\bibitem{Co:2019wyp}
R.~T. Co and K.~Harigaya, \emph{{Axiogenesis}},
  \href{https://doi.org/10.1103/PhysRevLett.124.111602}{\emph{Phys. Rev. Lett.}
  {\bfseries 124} (2020) 111602}
  [\href{https://arxiv.org/abs/1910.02080}{{\ttfamily 1910.02080}}].

\bibitem{Turner:1987bw}
M.~S. Turner and L.~M. Widrow, \emph{{Inflation Produced, Large Scale Magnetic
  Fields}}, \href{https://doi.org/10.1103/PhysRevD.37.2743}{\emph{Phys. Rev. D}
  {\bfseries 37} (1988) 2743}.

\bibitem{Garretson:1992vt}
W.~D. Garretson, G.~B. Field and S.~M. Carroll, \emph{{Primordial magnetic
  fields from pseudoGoldstone bosons}},
  \href{https://doi.org/10.1103/PhysRevD.46.5346}{\emph{Phys. Rev. D}
  {\bfseries 46} (1992) 5346}
  [\href{https://arxiv.org/abs/hep-ph/9209238}{{\ttfamily hep-ph/9209238}}].

\bibitem{Anber:2006xt}
M.~M. Anber and L.~Sorbo, \emph{{N-flationary magnetic fields}},
  \href{https://doi.org/10.1088/1475-7516/2006/10/018}{\emph{JCAP} {\bfseries
  10} (2006) 018} [\href{https://arxiv.org/abs/astro-ph/0606534}{{\ttfamily
  astro-ph/0606534}}].

\bibitem{Kamada:2018tcs}
K.~Kamada, \emph{{Return of grand unified theory baryogenesis: Source of
  helical hypermagnetic fields for the baryon asymmetry of the universe}},
  \href{https://doi.org/10.1103/PhysRevD.97.103506}{\emph{Phys. Rev. D}
  {\bfseries 97} (2018) 103506}
  [\href{https://arxiv.org/abs/1802.03055}{{\ttfamily 1802.03055}}].

\bibitem{Domcke:2020quw}
V.~Domcke, K.~Kamada, K.~Mukaida, K.~Schmitz and M.~Yamada, \emph{{Wash-In
  Leptogenesis}},
  \href{https://doi.org/10.1103/PhysRevLett.126.201802}{\emph{Phys. Rev. Lett.}
  {\bfseries 126} (2021) 201802}
  [\href{https://arxiv.org/abs/2011.09347}{{\ttfamily 2011.09347}}].

\bibitem{Yoshimura:1978ex}
M.~Yoshimura, \emph{{Unified Gauge Theories and the Baryon Number of the
  Universe}}, \href{https://doi.org/10.1103/PhysRevLett.41.281}{\emph{Phys.
  Rev. Lett.} {\bfseries 41} (1978) 281}.

\bibitem{Ignatiev:1978uf}
A.~Y. Ignatiev, N.~V. Krasnikov, V.~A. Kuzmin and A.~N. Tavkhelidze,
  \emph{{Universal CP Noninvariant Superweak Interaction and Baryon Asymmetry
  of the Universe}},
  \href{https://doi.org/10.1016/0370-2693(78)90900-0}{\emph{Phys. Lett. B}
  {\bfseries 76} (1978) 436}.

\bibitem{Weinberg:1979bt}
S.~Weinberg, \emph{{Cosmological Production of Baryons}},
  \href{https://doi.org/10.1103/PhysRevLett.42.850}{\emph{Phys. Rev. Lett.}
  {\bfseries 42} (1979) 850}.

\bibitem{Akamatsu:2013pjd}
Y.~Akamatsu and N.~Yamamoto, \emph{{Chiral Plasma Instabilities}},
  \href{https://doi.org/10.1103/PhysRevLett.111.052002}{\emph{Phys. Rev. Lett.}
  {\bfseries 111} (2013) 052002}
  [\href{https://arxiv.org/abs/1302.2125}{{\ttfamily 1302.2125}}].

\bibitem{Martins:1995tg}
C.~J. A.~P. Martins and E.~P.~S. Shellard, \emph{{String evolution with
  friction}}, \href{https://doi.org/10.1103/PhysRevD.53.R575}{\emph{Phys. Rev.
  D} {\bfseries 53} (1996) 575}
  [\href{https://arxiv.org/abs/hep-ph/9507335}{{\ttfamily hep-ph/9507335}}].

\bibitem{Martins:1996jp}
C.~J. A.~P. Martins and E.~P.~S. Shellard, \emph{{Quantitative string
  evolution}}, \href{https://doi.org/10.1103/PhysRevD.54.2535}{\emph{Phys. Rev.
  D} {\bfseries 54} (1996) 2535}
  [\href{https://arxiv.org/abs/hep-ph/9602271}{{\ttfamily hep-ph/9602271}}].

\bibitem{Martins:2000cs}
C.~J. A.~P. Martins and E.~P.~S. Shellard, \emph{{Extending the velocity
  dependent one scale string evolution model}},
  \href{https://doi.org/10.1103/PhysRevD.65.043514}{\emph{Phys. Rev. D}
  {\bfseries 65} (2002) 043514}
  [\href{https://arxiv.org/abs/hep-ph/0003298}{{\ttfamily hep-ph/0003298}}].

\bibitem{Vilenkin:1981bx}
A.~Vilenkin, \emph{{Gravitational radiation from cosmic strings}},
  \href{https://doi.org/10.1016/0370-2693(81)91144-8}{\emph{Phys. Lett. B}
  {\bfseries 107} (1981) 47}.

\bibitem{Turok:1984cn}
N.~Turok, \emph{{Grand Unified Strings and Galaxy Formation}},
  \href{https://doi.org/10.1016/0550-3213(84)90407-3}{\emph{Nucl. Phys. B}
  {\bfseries 242} (1984) 520}.

\bibitem{Quashnock:1990wv}
J.~M. Quashnock and D.~N. Spergel, \emph{{Gravitational Selfinteractions of
  Cosmic Strings}}, \href{https://doi.org/10.1103/PhysRevD.42.2505}{\emph{Phys.
  Rev. D} {\bfseries 42} (1990) 2505}.

\bibitem{Blanco-Pillado:2013qja}
J.~J. Blanco-Pillado, K.~D. Olum and B.~Shlaer, \emph{{The number of cosmic
  string loops}}, \href{https://doi.org/10.1103/PhysRevD.89.023512}{\emph{Phys.
  Rev. D} {\bfseries 89} (2014) 023512}
  [\href{https://arxiv.org/abs/1309.6637}{{\ttfamily 1309.6637}}].

\bibitem{Blanco-Pillado:2017oxo}
J.~J. Blanco-Pillado and K.~D. Olum, \emph{{Stochastic gravitational wave
  background from smoothed cosmic string loops}},
  \href{https://doi.org/10.1103/PhysRevD.96.104046}{\emph{Phys. Rev. D}
  {\bfseries 96} (2017) 104046}
  [\href{https://arxiv.org/abs/1709.02693}{{\ttfamily 1709.02693}}].

\bibitem{Blasi:2020mfx}
S.~Blasi, V.~Brdar and K.~Schmitz, \emph{{Has NANOGrav found first evidence for
  cosmic strings?}},
  \href{https://doi.org/10.1103/PhysRevLett.126.041305}{\emph{Phys. Rev. Lett.}
  {\bfseries 126} (2021) 041305}
  [\href{https://arxiv.org/abs/2009.06607}{{\ttfamily 2009.06607}}].

\bibitem{Blasi:2020wpy}
S.~Blasi, V.~Brdar and K.~Schmitz, \emph{{Fingerprint of low-scale leptogenesis
  in the primordial gravitational-wave spectrum}},
  \href{https://doi.org/10.1103/PhysRevResearch.2.043321}{\emph{Phys. Rev.
  Res.} {\bfseries 2} (2020) 043321}
  [\href{https://arxiv.org/abs/2004.02889}{{\ttfamily 2004.02889}}].

\end{thebibliography}\endgroup
\end{document}